\documentclass{fac}
\usepackage{graphicx}
\usepackage{amssymb}
\usepackage{amsfonts}
\usepackage{amsmath}
\usepackage{dsfont}

\ifprodtf \else \usepackage{latexsym}\fi

\newcommand\black{\ensuremath{\blacktriangleright}}
\newcommand\white{\ensuremath{\vartriangleright}}

\newif\ifamsfontsloaded
\ifprodtf
  \newcommand\whbl{\white\kern-.1em--\kern-.1em\black}
  \newcommand\blwh{\black\kern-.1em--\kern-.1em\white}
  \newcommand\blbl{\black\kern-.1em--\kern-.1em\black}
  \newcommand\whwh{\white\kern-.1em--\kern-.1em\white}
  \amsfontsloadedtrue
\else
  \checkfont{msam10}
  \iffontfound
    \IfFileExists{amssymb.sty}
      {\usepackage{amssymb}\amsfontsloadedtrue
       \newcommand\whbl{\white\kern-.125em--\kern-.125em\black}%
       \newcommand\blwh{\black\kern-.125em--\kern-.125em\white}%
       \newcommand\blbl{\black\kern-.125em--\kern-.125em\black}%
       \newcommand\whwh{\white\kern-.125em--\kern-.125em\white}}
      {}
  \fi
\fi

\title[A Task Allocation Schema Based on Response Time Optimization in Cloud Computing]
      {A Task Allocation Schema Based on Response Time Optimization in Cloud Computing}

\author[Yong Wang]
    {Kai Li\\
    Yong Wang\\
    Meilin Liu\\
     School of Computer Science and Technology,\\
     Beijing University of Technology, Beijing, China\\
     }

\correspond{Yong Wang, Pingleyuan 100, Chaoyang District, Beijing, China.
            e-mail: wangy@bjut.edu.cn}

\pubyear{2011}
\pagerange{\pageref{firstpage}--\pageref{lastpage}}

\begin{document}
\label{firstpage}

\makecorrespond

\maketitle

\begin{abstract}
Cloud computing is a newly emerging distributed computing which is evolved from Grid computing. Task scheduling is the core research of cloud computing which studies how to allocate the tasks among the physical nodes so that the tasks can get a balanced allocation or each task's execution cost decreases to the minimum or the overall system performance is optimal. Unlike the previous task slices' sequential execution of an independent task in the model of which the target is processing time, we build a model that targets at the response time, in which the task slices are executed in parallel. Then we give its solution with a method based on an improved adjusting entropy function. At last, we design a new task scheduling algorithm. Experimental results show that the response time of our proposed algorithm is much lower than the game-theoretic algorithm and balanced scheduling algorithm and compared with the balanced scheduling algorithm, game-theoretic algorithm is not necessarily superior in parallel although its objective function value is better.
\end{abstract}


\section{Introduction}

As an Internet-based computing model of public participation, cloud computing\cite{1} is a large-scale distributed computing paradigm that is driven by economies of scale, in which a pool of abstracted, virtualized, dynamically-scalable, managed computing power, storage, platforms, and services are delivered on demand to external customers over the Internet. With the rapid development of the Internet, real-time data stream and connected devices' diverse development, SOA's adoption as well as the promotion of search service, social networks, mobile commerce and open collaboration's demands, cloud computing develops rapidly. Currently, Google, IBM, Amazon, Microsoft and other IT vendors are making great efforts to research and promote cloud computing.

The characteristics of cloud computing, such as abstract and virtualized resources, on demand request of the Internet users, isolation for the accesses of different users, dynamic scalability, security under the unsafe networks determine that cloud computing has its own technical architecture as a new distributed computing paradigm. These characteristics will greatly influence the technologies in cloud computing corresponding to those in traditional distributed computing and distributed systems, such as task scheduling technology.

Cloud computing not only provides various business applications or personal applications via the Internet, but also includes the integration of tradition high performance computing under the new environments, for example, Amazon's elastic compute cloud provides high performance computing for the Internet users. Under the large-scale demands for services, especially for the requirements of computing intensive services, how to allocate resources effectively, and how to handle service requests is particularly important for cloud computing platform. There is no doubt that each characteristics of cloud computing will make the task scheduling in cloud computing special, such as the nature and characteristics of powerfully parallel processing capabilities.

In this paper, we introduce a new task scheduling approach based on response time. This approach satisfies the special requirements in cloud computing to some extent and is also novel in traditional distributed computing. We assume that a task can be split into task slices and take the maximum of all task slices' execution time as the response time. When decomposing a task, each scheduler expects its response time minimum. Then we design a task scheduling algorithm. This scheduling method uses response time as the accordance of scheduling decision, and is suitable for cloud computing platforms.

In Section $2$, we introduce the related works briefly. We discuss the system structure of a cloud, and introduce a cloud computing system model for computing intensive requirements in
Section $3$. In Section $4$, we establish the mathematical model based on response time and design the task scheduling algorithm. In Section $5$, we do a number of detailed experiments which show that this scheduling algorithm is better than the game-theoretic algorithm and the so-called balanced scheduling algorithm. Finally, we conclude this paper in Section 6.

\section{Related Works}

In general, job allocation algorithms in distributed systems can be classified as static or dynamic \cite{2}. In static algorithms, job allocation decisions are made at compile time and remain constant during runtime. For example, in \cite{3}, Kim and Kameda proposed a simplified load balancing algorithm, which targets at the minimizing the overall mean job response time via adjusting the each node's load in a distributed computer system that consists of heterogeneous hosts, based on the single-point algorithm originally presented by Tantawi and Towsley. Grosu and Leung\cite{4} formulated a static load balancing problem in single class job distributed systems from the aspect of cooperative game among computers. Also, there exists several studies on static load balancing in multi-class job systems \cite{5,6}. In contrast, dynamic job allocation algorithms attempt to use the runtime state information to make more informative job allocation decisions. In \cite{7}, Delavar introduced a new scheduling algorithm for optimal scheduling of heterogeneous tasks on heterogeneous sources, according to Genetic Algorithm which can reach to better makespan and more efficiency. In \cite{8}, Fujimoto proposed a new algorithm RR that uses the criterion called total processor cycle consummation, which is the total number of instructions the grid could compute until the completion time of the schedule, regardless how the speed of each processor varies over time, the consumed computing power can be limited within $(1+m(\ln(m-1)+1)/n)$( $m$ represents the number of the processor, n represents the number of independent coarse-grained tasks with the same length)times the optimal one.

For balanced task scheduling, \cite{9,10,11} proposed some models and task scheduling algorithms in distributed system with the market model and game theory. \cite{12,13} introduced a balanced grid task scheduling model based on non-cooperative game. QoS-based grid job allocation problem is modeled as a cooperative game and the structure of the Nash bargaining solution is given in \cite{2}. In \cite{14}, Wei and Vasilakos presented a game theoretic method to schedule dependent computational cloud computing services with time and cost constrained, in which the tasks are divided into subtasks. The above works generally take the scheduler or job manager as the participant of the game, take the total execution time of tasks as the game optimization goals and give the proof of the existence of the Nash equilibrium solution and the solving Nash equilibrium solution algorithm, or model the task scheduling problem as a cooperative game and give the structure of the cooperative game solution.

In a cloud computing environment, the goal of task scheduling is to achieve the optimal scheduling of jobs submitted by users, and try to improve the overall throughput of the cloud computing system. In recent years, a lot of people have been studying the task scheduling problems in the cloud computing environment and made rich achievements. At present, task scheduling algorithms at home and broad mainly base on the earliest completion time ,quality of service, load balancing, economic principles and so on. Job allocation algorithms can be classified as performance-centric, QoS-centric and economic principle-centric based on the different goals. Performance-centric task scheduling algorithms take the scheduling performance as the ultimate goal such as the shortest completion time, including Max-Min, Min-Min algorithm (e.g.,\cite{15,16}), genetic algorithm (e.g.\cite{17}), ant ant colony algorithm (e.g.,\cite{18}). QoS-centric task scheduling algorithms have been studied widely. In \cite{19}, Chanhan and Joshi selected resources based on the weighted average execution time, regarded the network bandwidth as QoS attributes and divided the tasks into two categories , high QoS and low QoS, and high QoS tasks had the priority to be scheduled. In \cite{20}, Xu and Wang developed a scheduling strategy for multiple workflows with different QoS requirements. In addition, \cite{21,22} proposed some task scheduling algorithms from the view of economic principle.

Unlike research works[12-13] directly related to this article in which the task slices are actually executed sequentially, the task slices generated by the scheduler are executed in parallel here.

According to the classification of static algorithms or dynamic ones for task scheduling algorithms, our algorithm can be classified into a semi-dynamic algorithm, just because this algorithm utilizes the states and capacity of computing nodes in a statistic way.

This paper has two main contributions as follows:
\begin{enumerate}
  \item Based on the cloud computing system model, we establish the mathematical model based on the response time.
  \item In view of difficulty of the above mathematical model for solving optimization problems, based on adjusting maximum entropy method, we give an approximate solution of the optimization model and design a new task scheduling algorithm. From the experiments, we can see that this scheduling algorithm has better optimal results than the game-theoretic algorithm and balanced scheduling algorithm.
\end{enumerate}

\section{System Model for A Cloud System}

\begin{figure}
 \begin{center}
  \includegraphics[width=8cm,height=8cm]{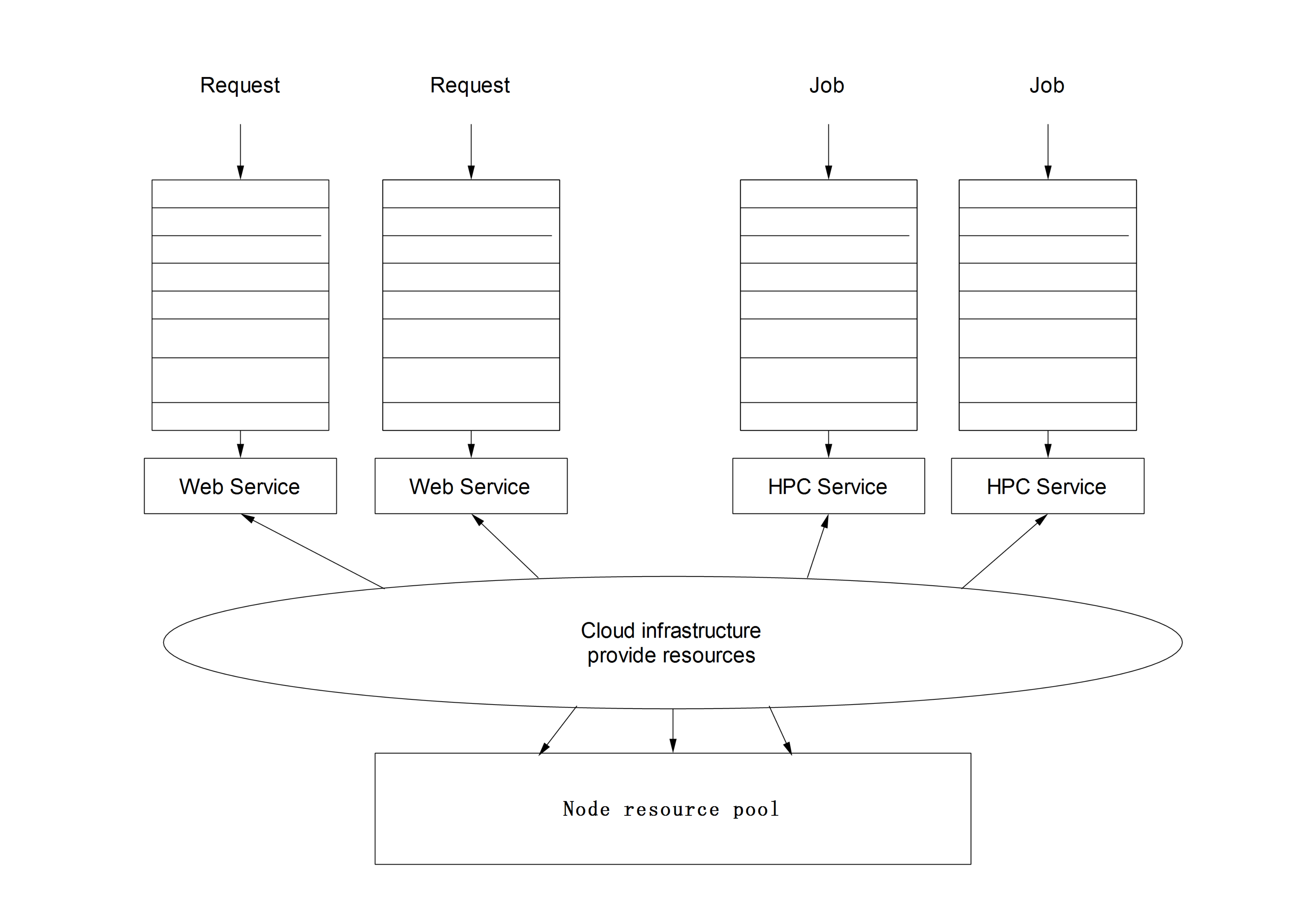}
  \caption{Cloud System Simulation}
  \label{Fig.1}
 \end{center}
\end{figure}

A variety of services in the cloud computing platform can be roughly grouped into two categories: data-compute-intensive services and interaction-intensive services \cite{23}. The former have a higher complexity and requires a higher computing power; the latter can be classified as general Web Services which needs higher real-time requirements. Fig.\ref{Fig.1} is a cloud system simulation based on the two type of services. Firstly, the service request will be classified to the Web Service queue or HPC (High performance computing) service queue according to its type, and then calls cloud resources depending on the type of service.

\begin{figure}
 \begin{center}
  \includegraphics[width=8cm,height=8cm]{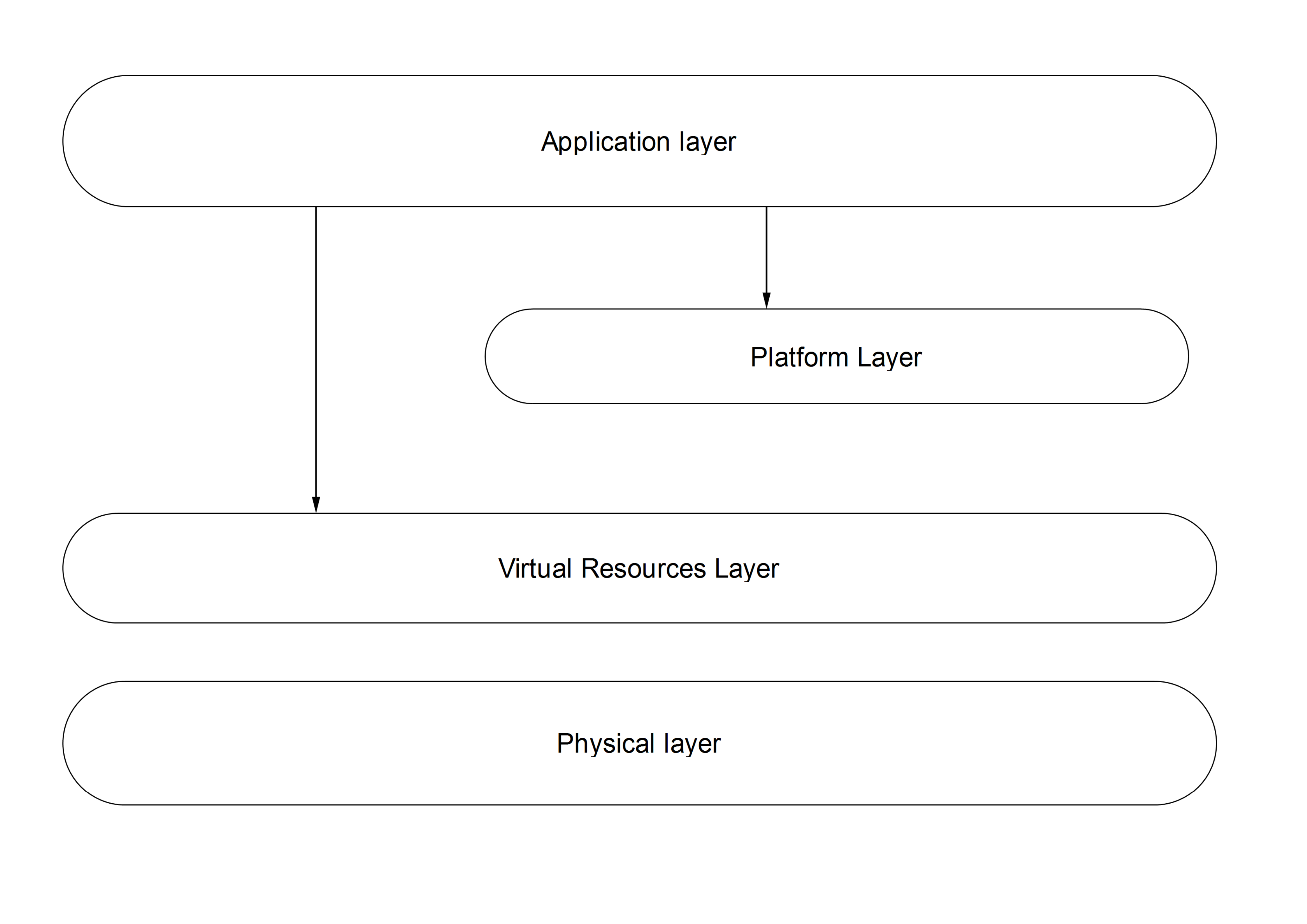}
 \caption{Cloud System Architecture}
 \label{Fig.2}
 \end{center}
\end{figure}

It is well known that the architecture of a cloud system roughly includes application layer, platform layer, virtual resource layer and physical layer \cite{1} shown as Fig.\ref{Fig.2}. The application layer accepts the requests of the Internet users through various applications, then deliver the requests to the platform layer. The platform layer usually contains a task scheduler to split the user tasks into pieces and deliver these task pieces onto different virtual resources. There are mappings from virtual resources to physical resources using the so-called virtualization technology.

In addition, data-compute-intensive services are usually very complex, and for these service requests, a cloud system should give full play to their own advantages of task parallel processing and the following three assumptions should be reasonable:

\begin{enumerate}
  \item In the cloud computing system model, the scheduler can split the task into task slices. Because the configuration of a node is controllable within a mesh range, we can assume that all the computing nodes can handle the task slices;
  \item The internal processing cost of the scheduler can be ignored, namely, we can assume that the task slices' transmission cost on the network and execution cost on the computing nodes are the key consideration of the task execution cost;
  \item In the cloud computing system, node resources that provide service may be subjected to the M/G/1 queueing system.
\end{enumerate}

\begin{figure}
 \begin{center}
  \includegraphics[width=8cm,height=8cm]{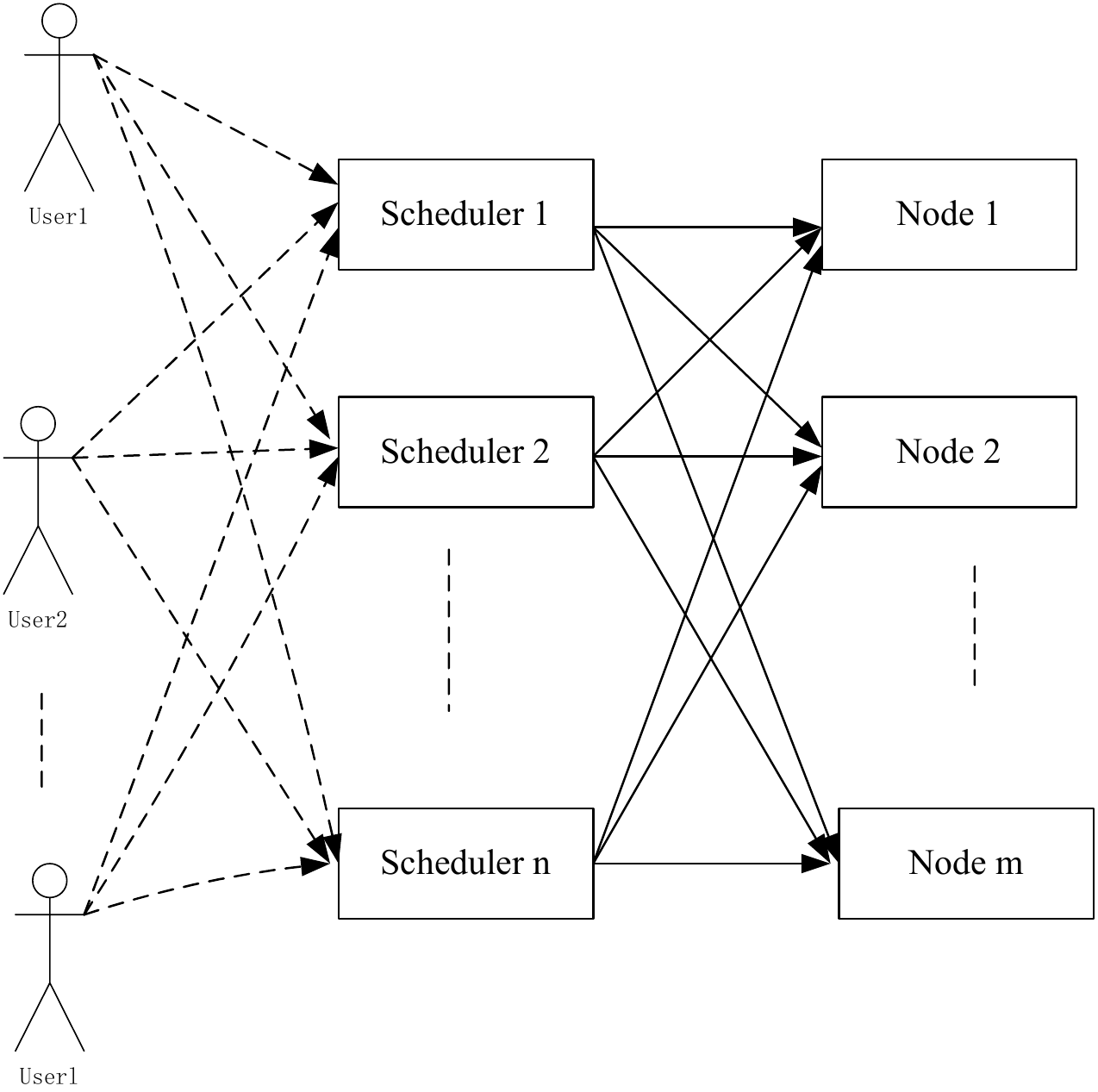}
 \caption{System Model}
 \label{Fig.3}
 \end{center}
\end{figure}

Conjunction with Fig.\ref{Fig.1} and Fig.\ref{Fig.2}, the system model of a cloud system for compute-intensive requirements shown in Fig.\ref{Fig.3} is reasonable. In this system model, the number of computing nodes that can provide service is assumed to be $m$, the number of users that generate service request is $l$, and the number of schedulers that allocates resource to implement the service request is $n$.

\begin{itemize}
  \item User: a user generates a request(task) to a scheduler and user $k$ is assumed to generate jobs with average rate $\beta_k$(jobs per second ) according to a Poisson process independently. Jobs are then sent by the user to a scheduler that dispatches them to the computing nodes.
  \item Scheduler: a scheduler receives job from a set of users and then assigns them to computing nodes in the cloud computing system. According to the first assumption, the task decomposition cost is negligible.
  \item Task slice: depending on the number of computing nodes, scheduler $i$ dispatches the users' tasks into $m$ task slices, $a_{ij}$ is the ratio of scheduler $i$ assigns one task to the computing node $j$ which satisfies the following constraint:
      \begin{equation}
           a_{ij}\geq0 \quad and \quad \sum_{j=1}^m a_{ij}=1
      \end{equation}
  \item Computing node: the computing node executes and processes task slices sent to it. According to the second assumption, the computing node has the ability to execute the general task slice. $\mu_j$is the average processing rate of jobs at computational node $j$, the execution time can be subjected to arbitrary distribution, and each computing node can be modeled as an M/G/1 queuing system.
      For stability, we have two constrains as follows:
  \begin{equation}
     \sum_{i=1}^n \lambda_i<\sum_{j=1}^m \mu_j
  \end{equation}

  \begin{equation}
     \sum_{i=1}^n \lambda_ia_{ij}<\mu_j
  \end{equation}

Where $\lambda_i$ is the average rate of jobs that scheduler $i$ issues. The first constrain  means that the average rate of jobs that all schedulers issues must not be faster than average rate of jobs that all computing nodes execute. The second constrain means that the average rate of jobs sent to computing node $j$ must not exceed the rate at which jobs can be executed by the computing node $j$.
\end{itemize}

\section{Mathematical Model to Optimize the Task Response Time}

\subsection{Objective Function}

Let $a_i=\{a_{i1},a_{i2},...,a_{im}\}$ represents scheduler $i$'s task slicing scheme on all cloud computing nodes , where $i=1,2,...,n$ and $a_{ij}$ is the task's ratio of jobs according which scheduler $i$ assigns jobs to computing node $j$. We define the maximum processing time among the task slices as this task's response time. Each scheduler expects that its task response time will be the shortest, thus, we can establish the objective function based on the task response time.
From the above assumption two, we can know that task slice's processing time contains two parts: the transmission time from a scheduler to a computing node and the execution time at the computing node.
Assume that the average length of all tasks is $b$ bits, the transmission delay time from scheduler $i$ to the computing node $j$ is $e_{ij}$, the communication bandwidth from scheduler $i$ to computing node $j$ is $c_{ij}$, the transmission time of task slice $a_{ij}$ can be calculated by formula 4.
\begin{equation}
    L_{ij}=e_{ij}+\frac{b\times a_{ij}}{c_{ij}}
\end{equation}
According to reference\cite{24}, the average service time for the M/G/1 queuing system is as follows:
\begin{equation}
    E(t)=\frac{1}{\mu}+\frac{\lambda(\sigma^{2}+\frac{1}{\mu^{2}})}{2(1-\frac{\lambda}{\mu})}
\end{equation}
Where $\sigma^{2}$ is the variance of the service time, $\mu$ is the average execution rate of the computing node.
Assume that the computing node's service time subjects to the negative exponential distribution, thus the average service time $F_{ij}$of task slice $a_{ij}$ on the computing node is shown as follows:
\begin{equation}
    F_{ij}=(\frac{1}{\mu_{j}}+\frac{\sum_{k=1}^{n}\lambda_{k}a_{kj}}{\mu_{j}(\mu_{j}-\sum_{k=1}^{n}\lambda_{k}a_{kj})})\times a_{ij}
\end{equation}
The execution time of task slice $a_{ij}$ is the sum of the transmission time $L_{ij}$ and the average service time $F_{ij}$ on the computing node $j$,namely:
\begin{equation}
    F_{ij}+L_{ij}=(\frac{1}{\mu_{j}}+\frac{\sum_{k=1}^{n}\lambda_{k}a_{kj}}{\mu_{j}(\mu_{j}-\sum_{k=1}^{n}\lambda_{k}a_{kj})})\times a_{ij}+e_{ij}+\frac{b\times a_{ij}}{c_{ij}}
\end{equation}
After the slices of a task are scheduled to the computing nodes, each slice is executed independently. the response time of this task is:
\begin{equation}
\begin{split}
    &FL_{i}(a_{i})=\max_{j=1}^{m}(F_{ij}+L_{ij})\\
    &=\max_{j=1}^{m}((\frac{1}{\mu_{j}}+\frac{\sum_{k=1}^{n}\lambda_{k}a_{kj}}{\mu_{j}(\mu_{j}-\sum_{k=1}^{n}\lambda_{k}a_{kj})})\times a_{ij}+e_{ij}+\frac{b\times a_{ij}}{c_{ij}})
\end{split}
\end{equation}
Each scheduler when scheduling the task, always expects a minimum response time, and therefore we can get the objective function of scheduler $i$ based on response time optimization:
\begin{equation}
   D_{i}=\min\max_{j=1}^{m}((\frac{1}{\mu_{j}}+\frac{\sum_{k=1}^{n}\lambda_{k}a_{kj}}{\mu_{j}(\mu_{j}-\sum_{k=1}^{n}\lambda_{k}a_{kj})})\times a_{ij}+e_{ij}+\frac{b\times a_{ij}}{c_{ij}})
\end{equation}

In order to facilitate the algorithm description, we introduce a new variable $\mu_{ji}$ ,shown in (10). $\mu_{ji}$ is defined as the computational  power of computing node $j$ that is available to scheduler $i$ and can be estimated for each computing node $j$:
\begin{equation}
 \mu_{ji}=\mu_{j}-\sum_{k=1,k\neq i}^{n}\lambda_{k}a_{kj}
\end{equation}
Using(10), (9) becomes:
\begin{equation}
   D_{i}=\min\max_{j=1}^{m}((1+\frac{\mu_{j}-\mu_{ji}+\lambda_{i}a_{ij}}{\mu_{ji}-\lambda_{i}a_{ij}})\times\frac{a_{ij}}{\mu_{j}}+e_{ij}+\frac{b\times a_{ij}}{c_{ij}})
\end{equation}
(7) is convex , which can be proven as follows:
\begin{equation}\nonumber
   \frac{\partial(F_{ij}+L_{ij})}{\partial a_{ij}}=\frac{\mu_{ji}}{(\mu_{ji}-\lambda_{i}a_{ij})^{2}}+\frac{b}{c_{ij}}>0
\end{equation}
\begin{equation}\nonumber
   \frac{\partial^{2}(F_{ij}+L_{ij})}{\partial a_{ij}^2}=\frac{2\lambda_{i}\mu_{ji}}{(\mu_{ji}-\lambda_{i}a_{ij})^{3}}>0
\end{equation}

\subsection{Adjusting Entropy Function Method}

When $m\geq2$, (11) is a non-differentiable optimization problem and it is difficult to get its solution. [25] introduces a maximum entropy method to solve the unconstrained minimax problem:
 \begin{equation}
 \min\max_{i=1}^{m}f_{i}(x)
 \end{equation}
 where $f_{i}(x)$ is a continuously differentiable function in $R^{n}$.

 The maximum entropy function is shown in (13).
 \begin{equation}
    FL^{p}(x)=\frac{1}{p}\ln\{\sum_{i=1}^{m}\exp(pf_{i}(x))\}
 \end{equation}

 When $p\rightarrow\infty$, $FL^{p}(x)$ uniformly converges to $$\max_{i=1}^{m}f_{i}(x)$$. Because this method simply increases the variable $p$, it is very difficult to program practically when $p$ is very large. In order to solve this problem, [26] introduces an improved entropy function shown in (14) and proposes an adjustable function algorithm.
 \begin{equation}
 F_{p}(x,\mu)=\frac{1}{p}\ln|\sum_{i=1}^{m}\mu_{i}\exp\{pf_{i}(x)\}|
\end{equation}
The variable parameter $\mu$  can be calculated by (15).
\begin{equation}
\mu_{i}^{(k+1)}=\frac{\mu_{i}^{k}\exp\{p^{(k)}f_{i}(x)\}}{\sum_{j=1}^{m}\mu_{j}^{k}\exp\{p^{(k)}f_{j}(x)\}}
\end{equation}
where $\mu_{i}^{(0)}=\{\frac{1}{m},\frac{1}{m},...,\frac{1}{m}\}$, $i=1,2,...,m$.

The adjustable function algorithm which is used to solve the minimax problem (12) is shown as follows:
\begin{itemize}
    \item Step $1$. Given a sufficiently large value $P>0$, $\varepsilon>0$, $p^{(0)}>0$, $\mu_{i}^{(0)}>0$, $r>1$, $k=0$, the initial point $x^{(0)}$;
    \item Step $2$. Solve $min F_{p^{(k)}}(x,\mu^{(k)})$ with $x^{(k)}$ as the starting point to get the new point $x^{(k+1)}$. if $\|x^{(k+1)}-x^{(k)}\|_{2}<\varepsilon$, stop, else go on;
    \item Step $3$. Calculate $\mu_{i}^{(k+1)}$ with (15), $i=1,2,...,m$. if $p^{(k)}<P$, $p^{(k+1)}=rp^{(k)}$, else $p^{(k+1)}=p^{(k)}$. $k=k+1$, goto step $2$;
\end{itemize}

This algorithm is very effective by adjusting $p$ and $\mu$, avoiding $p$ being too large.

\subsection{Task scheduling algorithm}

For the objective function (9), we can design an approximate algorithm on the basis of the above method as the following process: each scheduler calculates a best task slicing scheme $a_{i}$ that results in minimum $D_{i}$ in cycles until all schedulers' task slicing schema $a$ reaches an equilibrium or the cycles reach a certain value. In this paper, $f_{i}(x)$ is replaced by $F_{ij}+L_{ij}$.

On the basis of the above analysis, we obtain the task scheduling algorithm as follows:
\begin{itemize}
  \item Step $1$. System parameters initialization: Let $n$ represents the number of schedulers in the cloud computing system, $m$ represents the number of the computing nodes, $\lambda_i(0)$ is the average rate of jobs that scheduler $i$ issues, $\mu_j(0)$ is the average processing rate of jobs at computing node $j$, the transmission delay time from scheduler $i$ to computing node $j$ is $e_{ij}(0)$, the average data length of all tasks is $b(0)$ bits,  the communication bandwidth from scheduler $i$ to computing node $j$ is $c_{ij}(0)$ Kbps, where $i=1,2,...,n, j=1,2,...,m$. The task slice program of scheduler $i$ is initialized as: $a_i(0)=\{a_{i1}(0),a_{i2}(0),...,a_{im}(0)\}=\{\frac{1}{m},\frac{1}{m},...,\frac{1}{m}\}$, $maxCyle(0)$ is the maximum cycle number, the current cycle $currentCycle$ is $1$ , the maximum adjusting value is $P(0)$, $formerA=latterA=a(0)$, the initial value of $formerA$ and $latterA$ 's norm difference $diffA$ is $1$, the error precision is $\varepsilon(0)=10^{-4}$;
  \item Step $2$. Based on the above initial values, calculate $\mu_{ji}$ by (10);
  \item Step $3$. if $diffA>\varepsilon$ and $currentCycle<maxCycle$, then the task scheduling program is calculated and the cycle ends; otherwise, repeat the following step 4 to step 5;
  \item Step $4$: Perform the following steps from $i=1$ to $n$;
  \begin{itemize}
    \item Step $4.1$. Let $x0=latterA_{i}$, where $latterA_{i}$ represents the $i$th row of $latterA$, $\mu^{(0)}=\{\frac{1}{m},\frac{1}{m},,...,\frac{1}{m}\}$, $p=10$, $r=10$, $k=0$, $formerX=latterX=x0$;
    \item Step $4.2$. Solve $min F_{p^{(k)}}(x,\mu^{(k)})$ with $latterX$ as $x$'s starting point to get the new point $x$;
    \item Step $4.3$. Let $formerX=latterX$, $latterX=x$. if $\|formerX-latterX\|_{2}<\varepsilon$, goto step $4.5$, else go on;
    \item Step $4.4$. Calculate $\mu_{i}^{(k+1)}$ with (15), $i=1,2,...,m$, $k=k+1$. if $p^{(k)}<P$, $p^{(k+1)}=rp^{(k)}$, else $p^{(k+1)}=p^{(k)}$. Goto step $4.2$;
    \item Step $4.5$. $latterA_{i}=latterX$, calculate $\mu_{ji}$ with the new $latterA$;
    \end{itemize}
    \item Step $5$. Let $diffA=\|latterA-formerA\|_{2}$, $formerA=latterA$, goto step $3$.
\end{itemize}

The above task scheduling algorithm determines the task slicing program according to which each scheduler should allocate jobs to each computing node in order to minimize all schedulers' response time.

\section{Experiments}

In this section, we do a series of experiments to analyze the merits and demerits of this schema by comparison with other algorithms, such as game-theoretic algorithms \cite{13} and the balanced scheduling algorithm \cite{27}. The proposed algorithm is labeled as PS. The game-theoretic algorithm uses the game theory to design a new task scheduling algorithm with goal of completion time, which is labeled GS .The balanced scheduling algorithm allocates tasks to processors in proportion to its computing power(task processing rate), which is labeled BS; the faster computing node are sent more tasks by the schedulers. The proportion of tasks sent to the computing nodes is given by the following:
\begin{equation}
 a_{ij}=\frac{\mu_{ji}}{\sum_{j=1}^{m}\mu_{ji}}
\end{equation}
Let $\phi_{i}$ represents the relative job arrival rate of scheduler $i$, the average arrival rate of scheduler $i$ is calculated under the below formula.
\begin{equation}
   \lambda_{i}=\phi_{i}\cdot\rho\cdot\sum_{j=1}^{m}\mu_{j}
\end{equation}
Where $\rho$ is the required overall average system load.

\subsection{Objective Function Value and Response Time Value}

In a cloud computing system, the processing abilities of the computing nodes may be have little difference or not. Considering these two cases, we do two sets of experiments. In each case, we compare the objective function values and response time values of the above three different task scheduling algorithms.
Firstly, we set the overall average system load to 50 percent($\rho=0.5$), the average length of all tasks is $1$Mbits($b=1$Mbits), the average transmission delay time from scheduler $i$ to computing node $j$ is 0.5second($e_{ij}=0.5$s), the communication bandwidth from scheduler $i$ to computing node $j$ is $100$Kbps. At the same time, we assume that this cloud computing system has $n=7$ schedulers and $m=8$ computing nodes. The relative job arrival rate of each scheduler is shown at Table \ref{Table.1}.
  \begin{center}
  \begin{table}
  \caption{Relative Job Arrival Rate of Each Scheduler}
  \label{Table.1}
  \begin{tabular}{|c|c|c|c|c|}
    \hline
    scheduler & 1 & 2-5 & 6 & 7 \\
    \hline
    Relative job arrival rate & 0.0035 & 0.01 & 0.006 & 0.005\\
    \hline
  \end{tabular}
  \end{table}
  \end{center}

In the first set of experiment, there are some computing nodes that have higher average processing rate than the others clearly and the average processing rate of each computing nodes is shown at Table \ref{table.2}.
  \begin{center}
  \begin{table}
  \caption{Average Task Processing Rate of the Computing Nodes}
  \label{table.2}
  \begin{tabular}{|c|c|c|c|c|}
    \hline
    Computing node & 1 & 2 & 3 & 4\\
    \hline
    Average Task Processing Rate & 0.28 & 0.22 & 0.19 & 0.23 \\
    \hline
    Computing node  & 5 & 6 & 7 & 8\\
    \hline
    Average Task Processing Rate  & 0.20 & 0.26 & 0.22 & 0.23 \\
    \hline
  \end{tabular}
  \end{table}
  \end{center}

\begin{figure}
 \begin{center}
  \includegraphics[width=8cm,height=8cm]{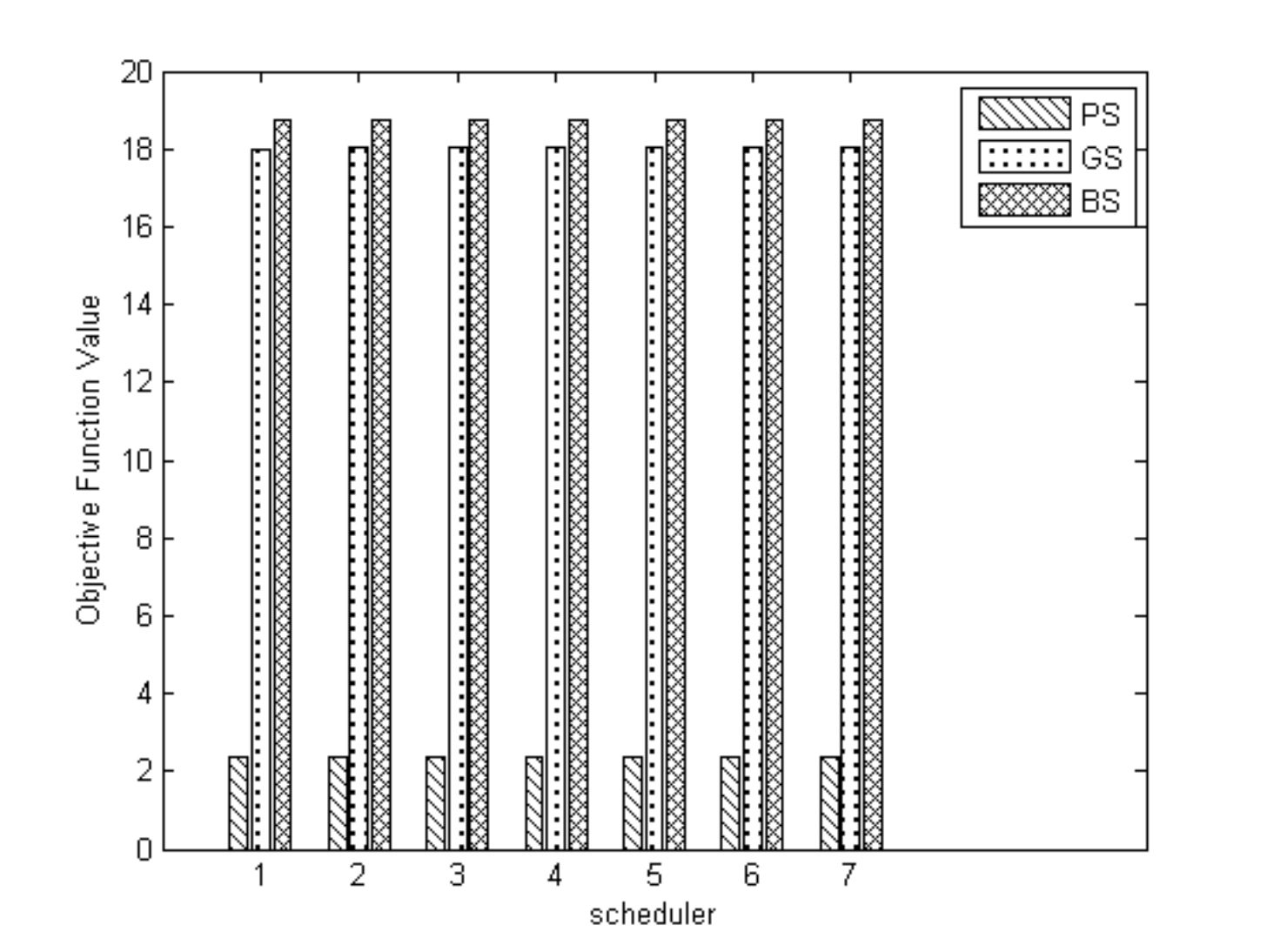}
 \caption{Objective Function Value of Each Scheduler}
 \label{Fig.4}
 \end{center}
\end{figure}

Under the above initial condition, the objective function value of each scheduler with different task scheduling algorithm is shown in Fig \ref{Fig.4} shows.

\begin{figure}
 \begin{center}
  \includegraphics[width=8cm,height=8cm]{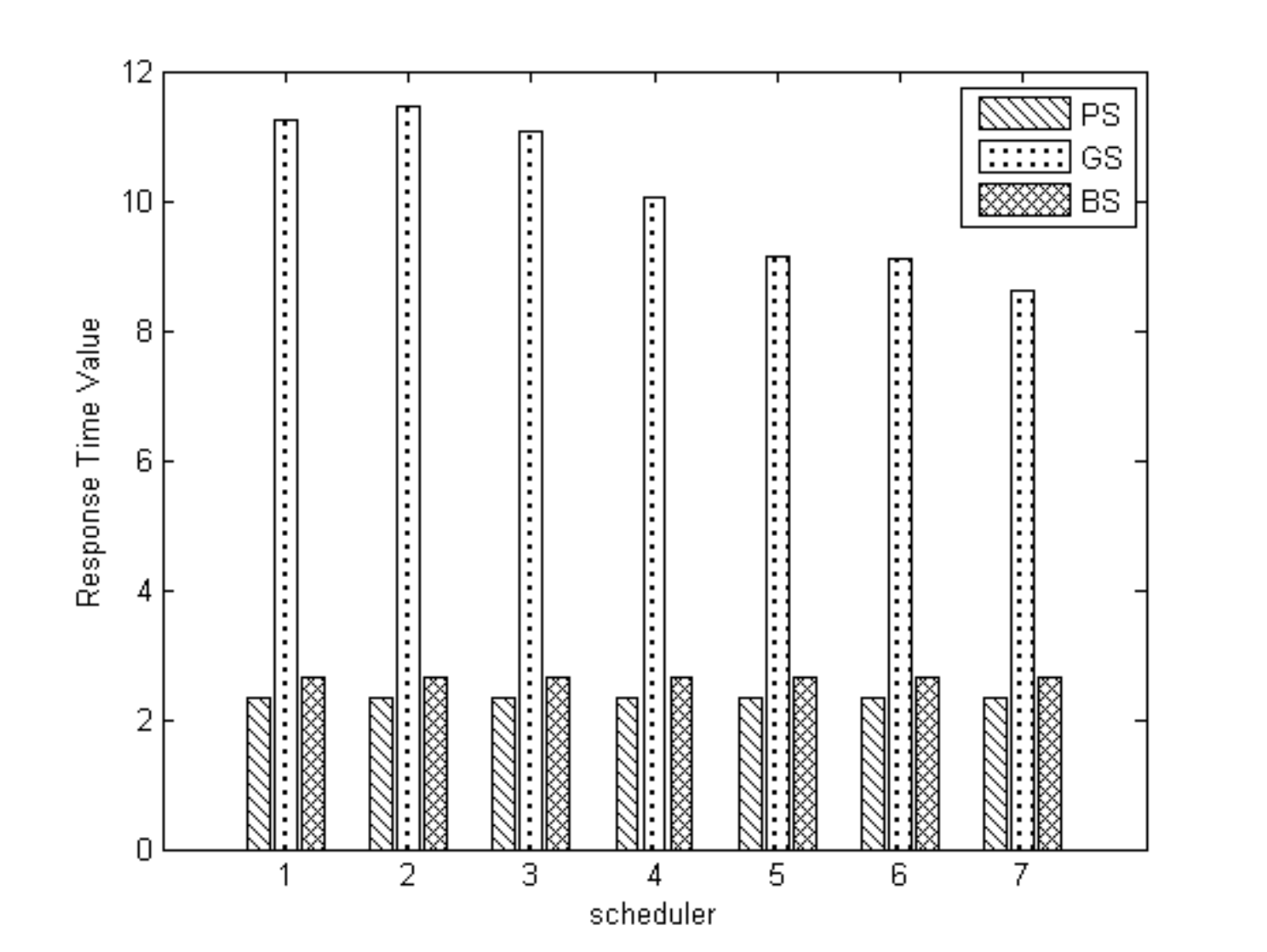}
 \caption{Response Time Value of Each Scheduler}
 \label{Fig.5}
 \end{center}
\end{figure}

As can be seen from Fig.\ref{Fig.4}, when the computing abilities of each computing node are different, the objective function value of each scheduler of our proposed task scheduling algorithm is significantly better than the other two algorithms. The game-theoretic algorithm is slightly better than the balanced scheduling algorithm. The objective function values among the schedulers are little different in each task scheduling algorithm.
In order to study the effects of GS and BS in the implementation of task slices executed in parallel, we apply these two algorithms to the response time function and get each scheduler's response time shown in Fig.\ref{Fig.5}.

From Fig.\ref{Fig.5}, each scheduler's response time of PS is slightly better than BS, and significantly better than GS. In addition, the response time values of PS and BS are more balanced than GS.

In the second set of experiment, the processing abilities of all computing nodes have little difference, which are shown in Table \ref{table.3}.
  \begin{center}
  \begin{table}
  \caption{Average Task Processing Rate of the Computing Nodes}
  \label{table.3}
  \begin{tabular}{|c|c|c|c|c|}
    \hline
    Computing node & 1 & 2 & 3 & 4\\
    \hline
    Average Task Processing Rate & 0.25 & 0.26 & 0.23 & 0.24 \\
    \hline
    Computing node  & 5 & 6 & 7 & 8\\
    \hline
    Average Task Processing Rate  & 0.22 & 0.25 & 0.22 & 0.23 \\
    \hline
  \end{tabular}
  \end{table}
  \end{center}

\begin{figure}
 \begin{center}
  \includegraphics[width=8cm,height=8cm]{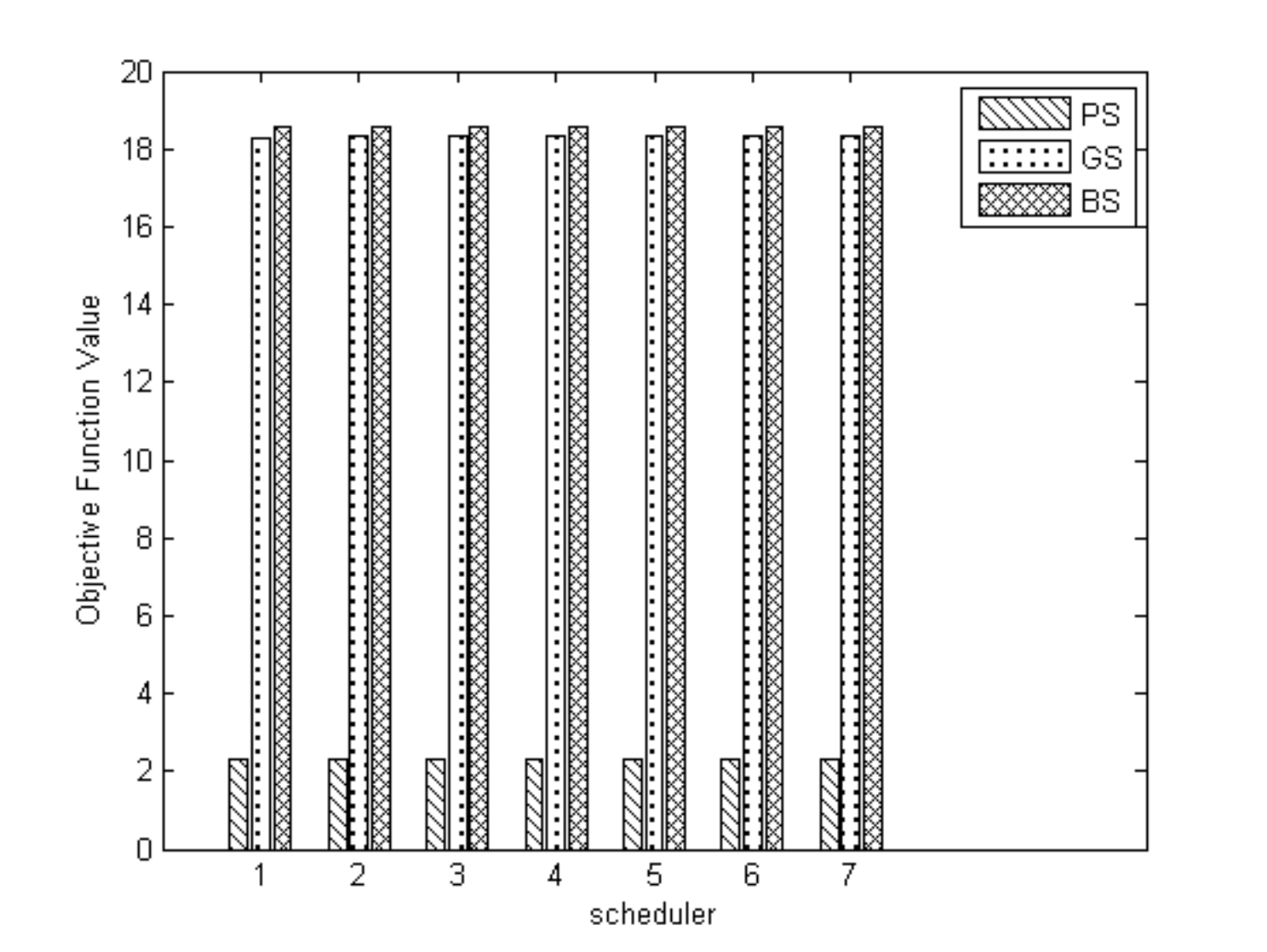}
 \caption{Objective Function Value of Each Scheduler}
 \label{Fig.6}
 \end{center}
\end{figure}

\begin{figure}
 \begin{center}
  \includegraphics[width=8cm,height=8cm]{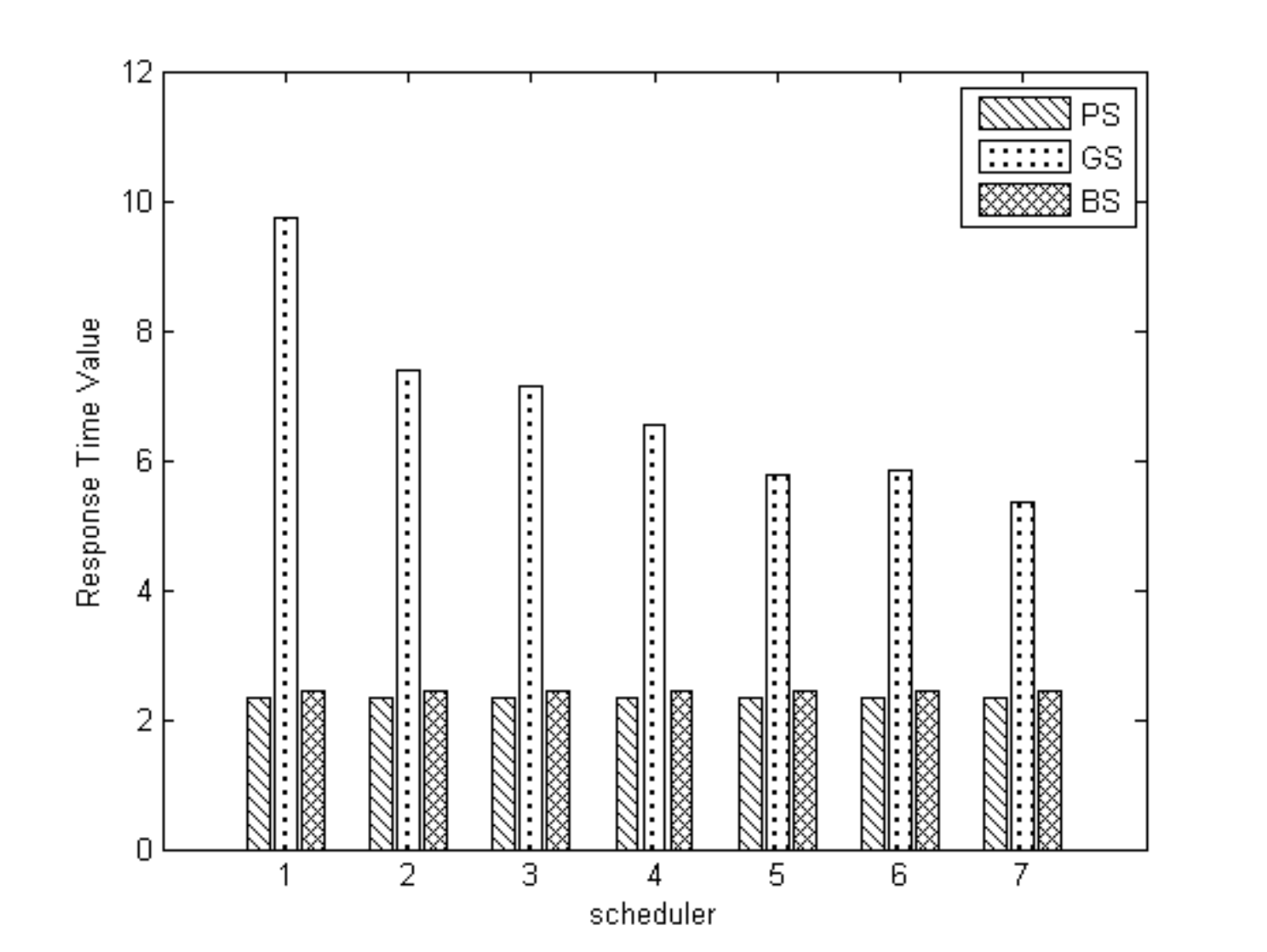}
 \caption{Response Time Value of Each Scheduler}
 \label{Fig.7}
 \end{center}
\end{figure}

Under the above initial condition, the objective function value and the response time value of each scheduler with different task scheduling algorithm are respectively shown in Fig.\ref{Fig.6} and Fig.\ref{Fig.7}.

From Fig.\ref{Fig.6} and Fig.\ref{Fig.7}, we can see that there are similar results whether the processing abilities of each computing node are balanced or not by comparison with the first experiment.
Combining these two sets of experiments, whether the cloud computing system provides computing nodes' processing abilities balanced or not, we can draw the following conclusions:

\textbf{Conclusion 1}: the response time values among the schedulers under PS and BS are balanced but for GS;

\textbf{Conclusion 2}: Both from response time value and the objective function value, the proposed task scheduling algorithm is superior to the other two algorithms;

\textbf{Conclusion 3}: Compared with BS, GS is not necessarily superior in parallel although its objective function value is better.

\subsection{Effects of System Load}

In this section, we compare the objective function values of the above three task scheduling algorithms when schedulers' arrival tasks increase. Here, we vary the average load of the system from $0.1$ to $0.9$. The remaining conditions are the same as the second set of experiment in the preceding experiments.

From the preceding experiments we can see that all scheduler's objective function values are similar under each task scheduling algorithms whether the processing abilities that the computing nodes provide are balanced or not. Therefore, we use the first scheduler's objective function value to compare the variation over system load in different algorithms. But the response time values among all schedulers are significantly different. In order to verify the merits of our proposed algorithm in response time, we compare the maximum of all schedulers' response time value under PS with the minimum of all schedulers' response time values under GS and BS. If the maximum value of PS is lower than the minimum of GS and BS, it is obvious that we can draw the conclusion that our algorithm is better than the others for the whole system.

\begin{figure}
 \begin{center}
  \includegraphics[width=8cm,height=8cm]{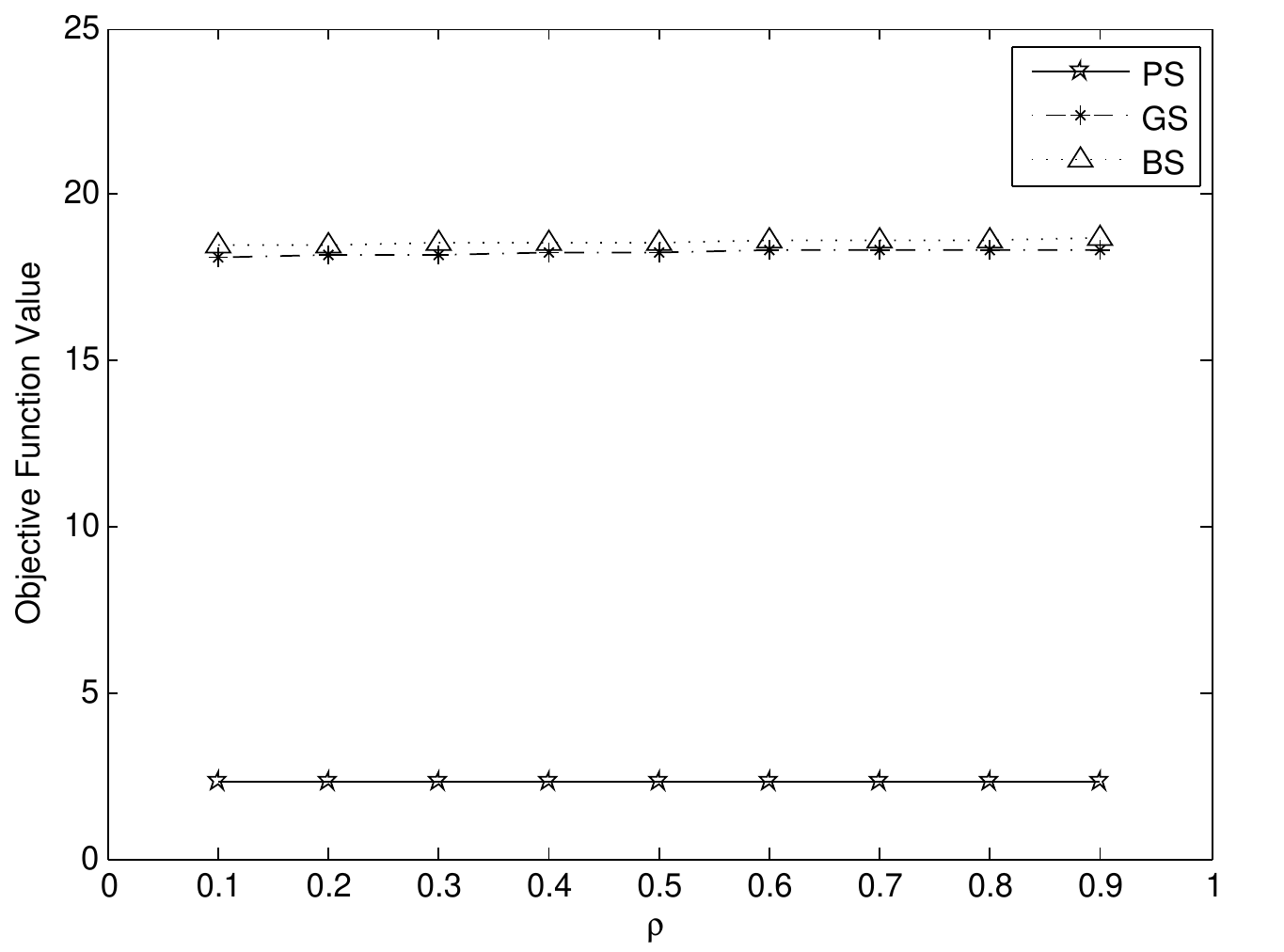}
 \caption{Objective Function Value versus System Load}
 \label{Fig.8}
 \end{center}
\end{figure}

Under the above initial conditions, we calculate the first scheduler's objective function values over the system load from $0.1$ to $0.9$, Fig.\ref{Fig.8} shows the results.

From Fig.\ref{Fig.8}, we can find that the objective function values under three algorithms is stable and the objective function value under PS is lower than GS and BS.

\begin{figure}
 \begin{center}
  \includegraphics[width=8cm,height=8cm]{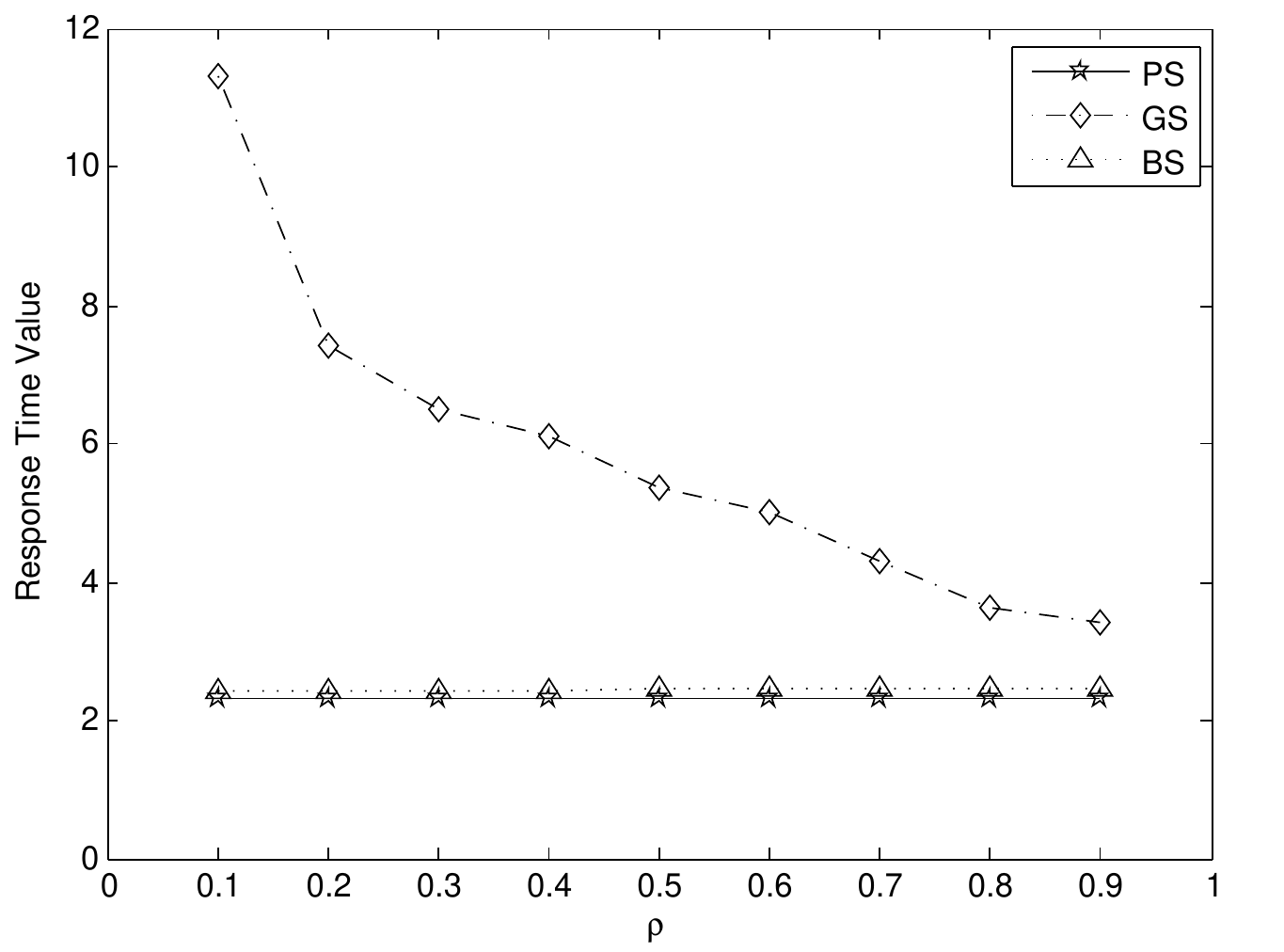}
 \caption{Response Time Value versus System Load}
 \label{Fig.8.1}
 \end{center}
\end{figure}

Fig.\ref{Fig.8.1} shows the maximum  of all schedulers' response time values under PS and the minimum of all schedulers' response time value under GS and BS with the system load varying from $0.1$ to $0.9$.

From Fig.\ref{Fig.8.1} we can find that: when the system load increases, although the minimum of all schedulers' response time value under GS decreases, PS is still better than GS and BS all the time.

With Fig.\ref{Fig.8} and Fig.\ref{Fig.8.1}, when the system load increases, our proposed algorithm is still better than the other two algorithms.

\subsection{Effects of System Size}

In this part of the experiment, system size mainly refers to the number of the schedulers and the number of computing nodes, and we also do two sets of experiments separately.

In the first set of experiment, we consider the effect of the number of the schedulers. We vary the number of the schedulers from $7$ to $15$, and the number of the computing nodes is $10$. Table \ref{table.4} shows ten computing nodes' average task processing rates and table \ref{table.5} shows fifteen schedulers'  relative job arrival rates. The remaining conditions are the same as the first experiment.

 \begin{center}
  \begin{table}
  \caption{Average Task Processing Rate of the Computing Nodes}
  \label{table.4}
  \begin{tabular}{|c|c|c|c|c|}
    \hline
    Computing node & 1 & 2 & 3 & 4-5 \\
    \hline
    Average Task Processing Rate & 0.25 & 0.26 & 0.23 & 0.23 \\
    \hline
    Computing node  & 6 & 7 & 8-9 & 10 \\
    \hline
    Average Task Processing Rate  & 0.21 & 0.24 & 0.24 & 0.22 \\
    \hline
  \end{tabular}
  \end{table}
  \end{center}

\begin{center}
  \begin{table}
  \caption{Relative Job Arrival Rate of Each Scheduler}
  \label{table.5}
  \begin{tabular}{|c|c|c|c|c|}
    \hline
    scheduler & 1 & 2-5 & 6 & 7 \\
    \hline
    Relative job arrival rate & 0.0035 & 0.01 & 0.006 & 0.005\\
    \hline
     scheduler & 8-10 & 11-13 & 14 & 15 \\
    \hline
    Relative job arrival rate & 0.003 & 0.002 & 0.0015 & 0.0015\\
    \hline
  \end{tabular}
  \end{table}
  \end{center}

\begin{figure}
 \begin{center}
  \includegraphics[width=8cm,height=8cm]{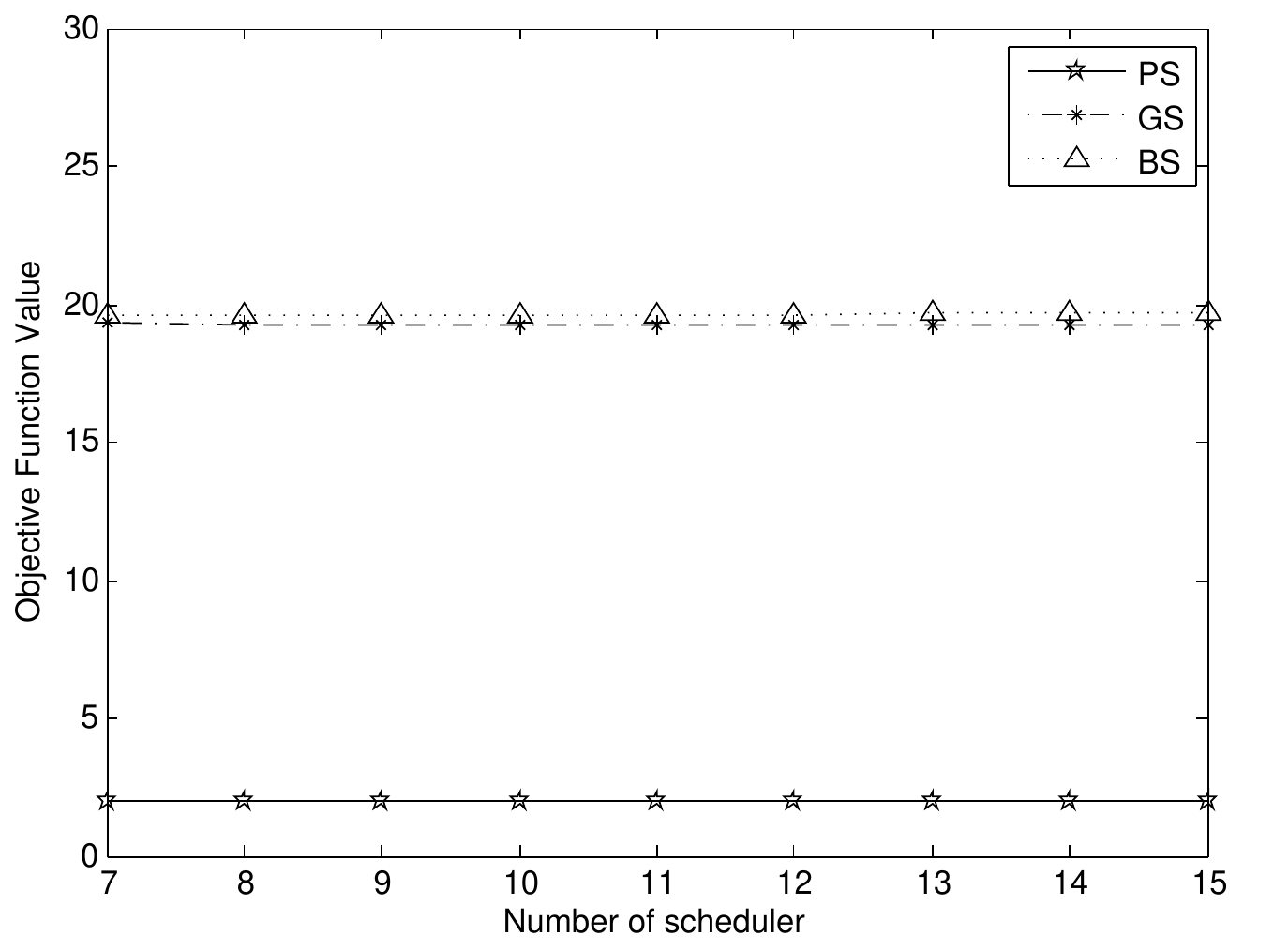}
 \caption{Objective Function Value versus the Number of Scheduler}
 \label{Fig.9}
 \end{center}
\end{figure}

\begin{figure}
 \begin{center}
  \includegraphics[width=8cm,height=8cm]{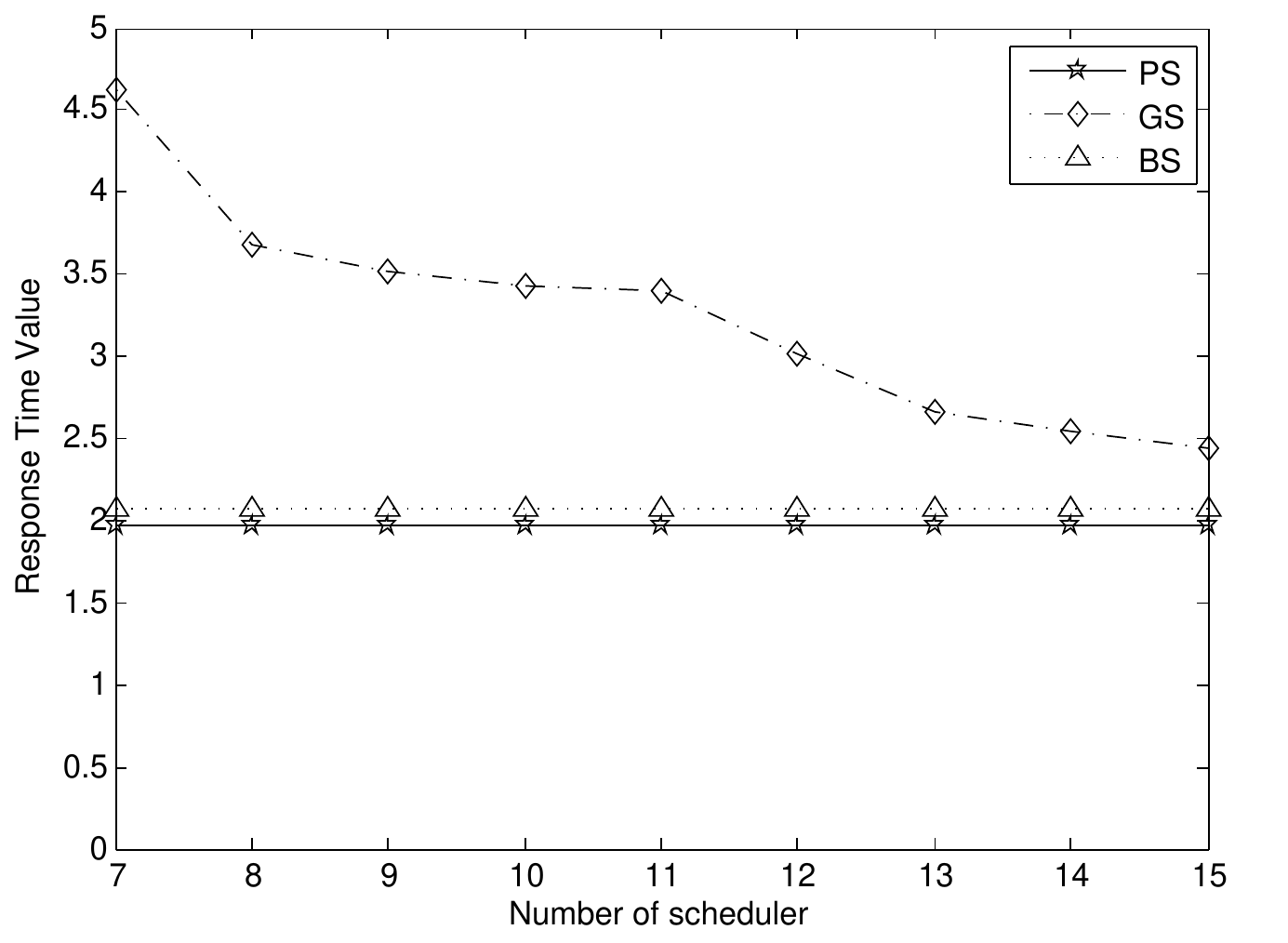}
 \caption{Response Time Value versus the Number of Scheduler}
 \label{Fig.9.1}
 \end{center}
\end{figure}

Under the above initial conditions, the first scheduler's objective function values are shown in Fig.\ref{Fig.9} and the maximum of all schedulers' response time value under PS with the minimum of all schedulers' response time values under GS and BS are shown in Fig.\ref{Fig.9.1} when the cloud computing system adds a scheduler each time

Fig.\ref{Fig.9} and Fig.\ref{Fig.9.1} show that our tasking scheduling algorithm is better than the other two algorithms when the number of schedulers increases.

In the second set of experiment, we investigate the effect of the number of computing nodes. We vary the number of computing nodes from $10$ to $15$ and set the number of the scheduler to $7$. The average task processing rate and the average  relative job arrival rate are listed in table \ref{table.6} and table\ref{table.7} separately. The remaining conditions are the same as the first experiment.
\begin{center}
  \begin{table}
  \caption{Average Task Processing Rate of the Computing Nodes}
  \label{table.6}
  \begin{tabular}{|c|c|c|c|c|}
    \hline
    Computing node & 1 & 2 & 3-5 & 6 \\
    \hline
    Average Task Processing Rate & 0.25 & 0.26 & 0.23 & 0.21 \\
    \hline
    Computing node  & 7-9 & 10-13 & 14 & 15 \\
    \hline
    Average Task Processing Rate  & 0.24 & 0.22 & 0.20 & 0.20\\
    \hline
  \end{tabular}
  \end{table}
  \end{center}

\begin{figure}
 \begin{center}
  \includegraphics[width=8cm,height=8cm]{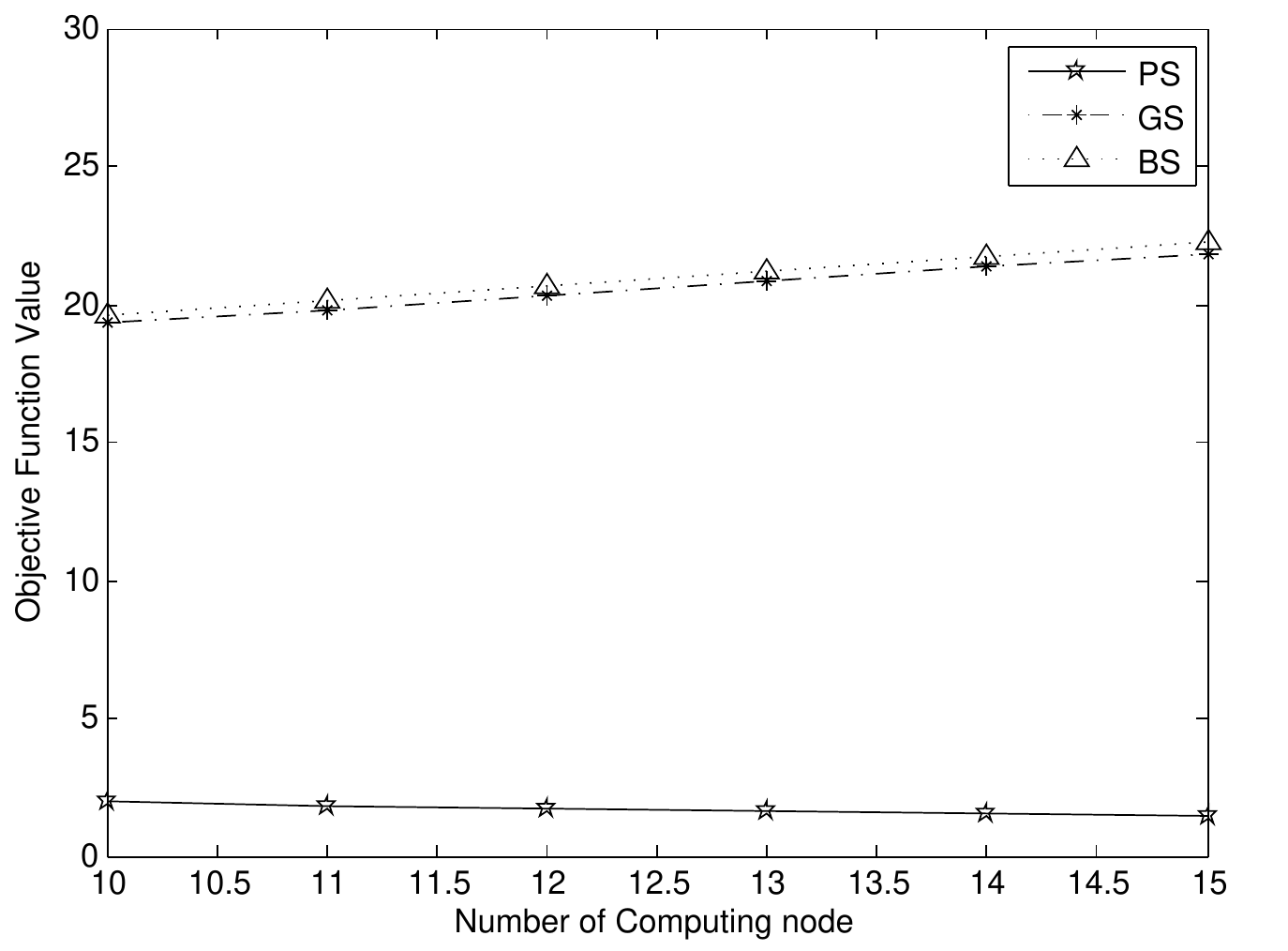}
 \caption{Objective Function Value versus the Number of Computing Node}
 \label{Fig.10}
 \end{center}
\end{figure}

\begin{figure}
 \begin{center}
  \includegraphics[width=8cm,height=8cm]{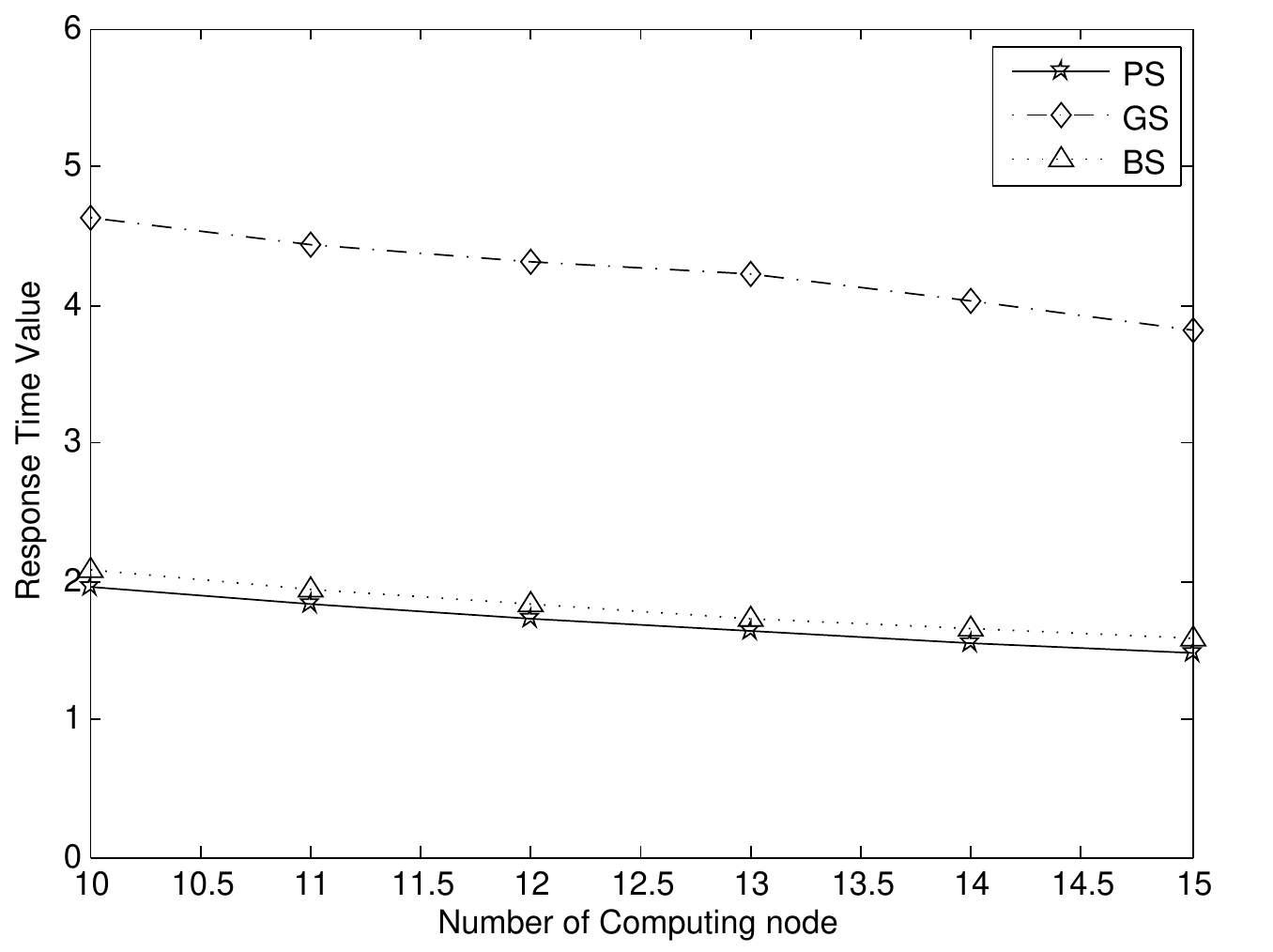}
 \caption{Response Time Value versus the Number of Computing Node}
 \label{Fig.10.1}
 \end{center}
\end{figure}

\begin{center}
  \begin{table}
  \caption{Relative Job Arrival Rate of Each Scheduler}
  \label{table.7}
  \begin{tabular}{|c|c|c|c|c|}
    \hline
    scheduler & 1 & 2-5 & 6 & 7 \\
    \hline
    Relative job arrival rate & 0.0035 & 0.01 & 0.006 & 0.005\\
    \hline
  \end{tabular}
  \end{table}
  \end{center}
Under the above conditions, the first scheduler's objective function values are shown in Fig.\ref{Fig.10} and the maximum of all schedulers' response time value under PS with the minimum of all schedulers' response time values under GS and BS are shown in Fig.\ref{Fig.10.1} when the cloud computing system adds a computing node each time.

As can be seen from Fig.\ref{Fig.10} and Fig.\ref{Fig.10.1}, our tasking scheduling algorithm is better than the other two algorithms when the number of computing nodes increases.

According to the above two sets of experiments, we can draw this conclusion: when the system size increases, PS is still better than the other two algorithms.

\subsection{Effects of Communication Bandwidth}

\begin{figure}
 \begin{center}
  \includegraphics[width=2.5cm,height=2.5cm]{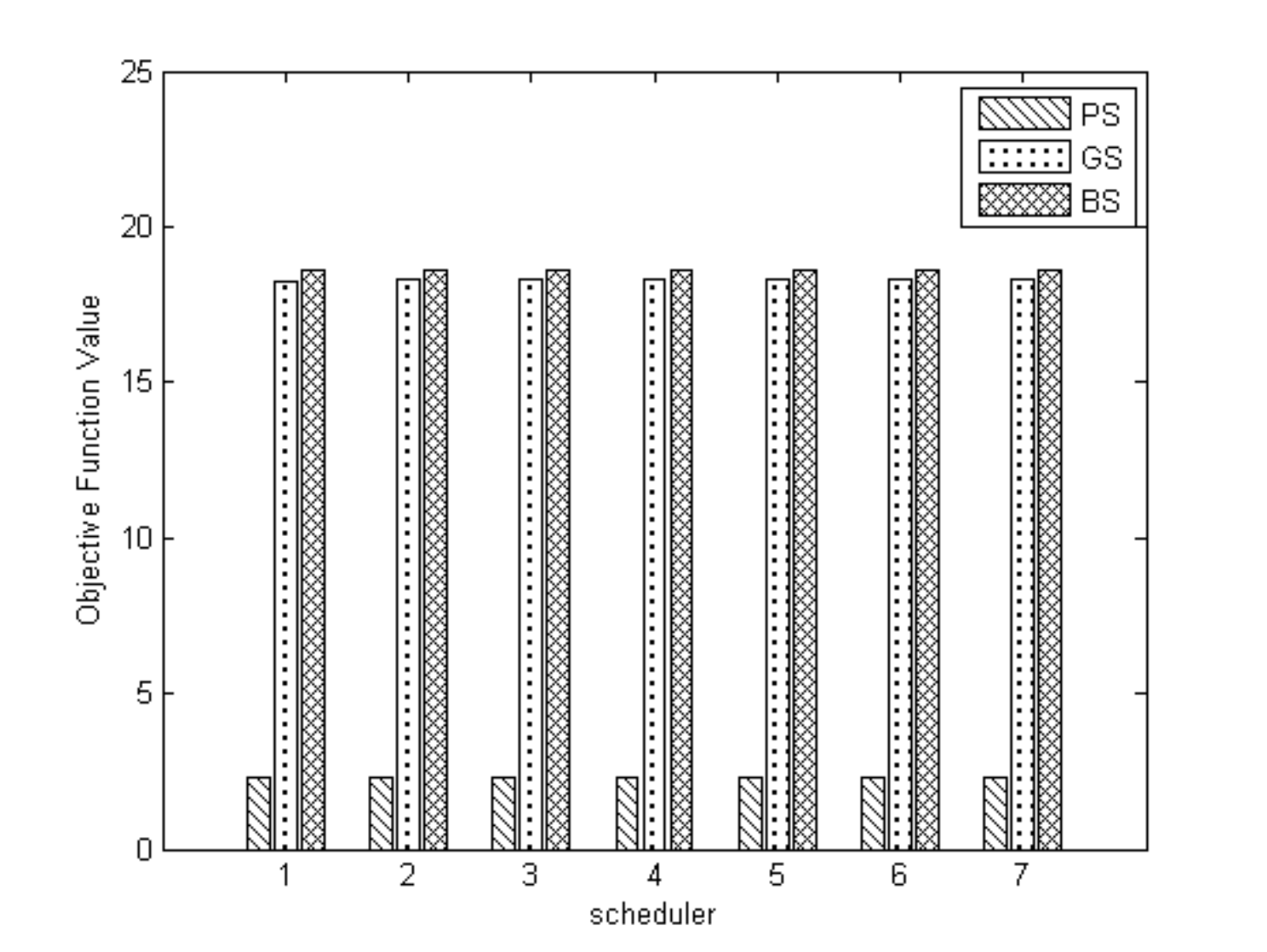}
  \includegraphics[width=2.5cm,height=2.5cm]{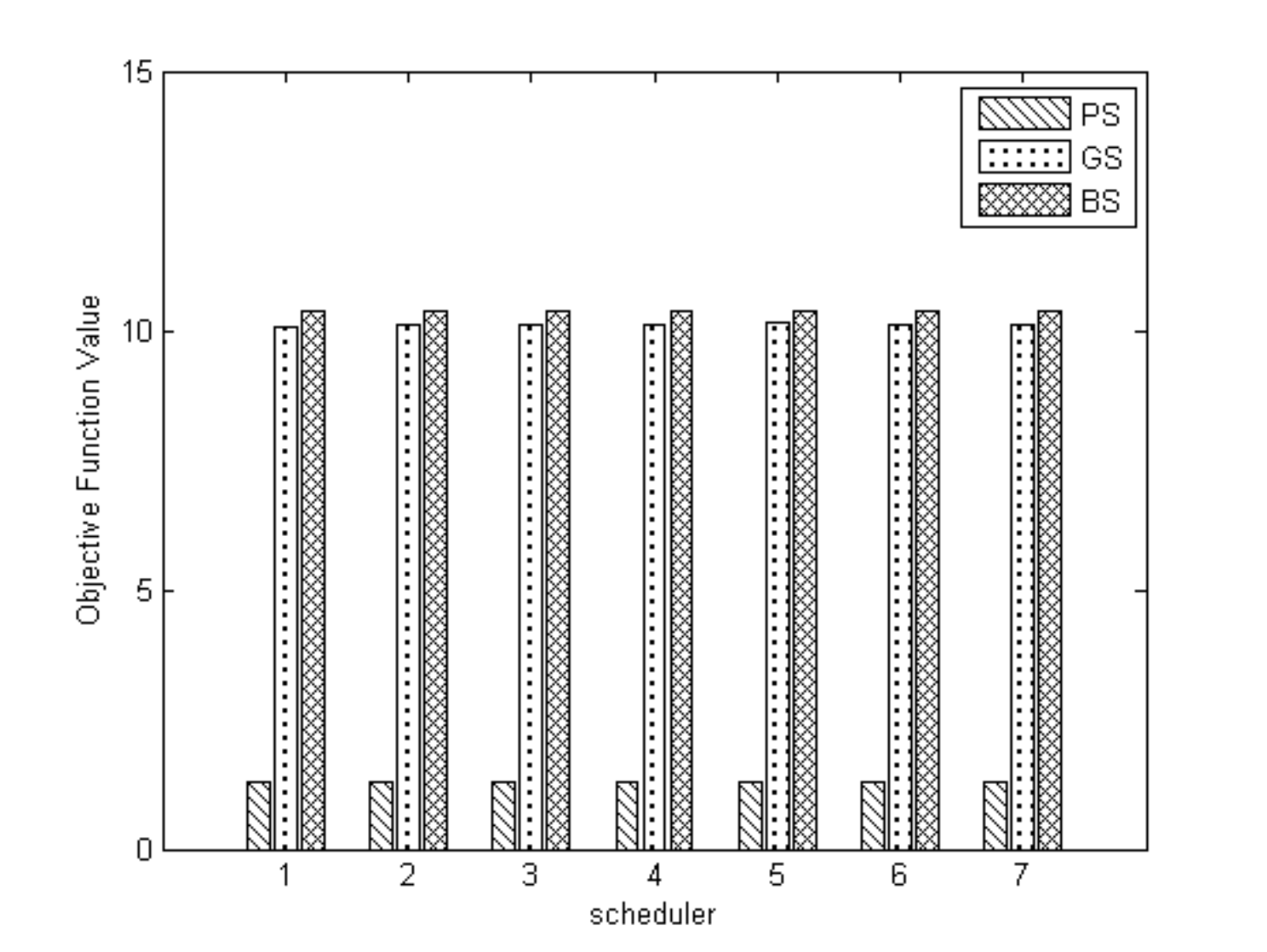}
  \includegraphics[width=2.5cm,height=2.5cm]{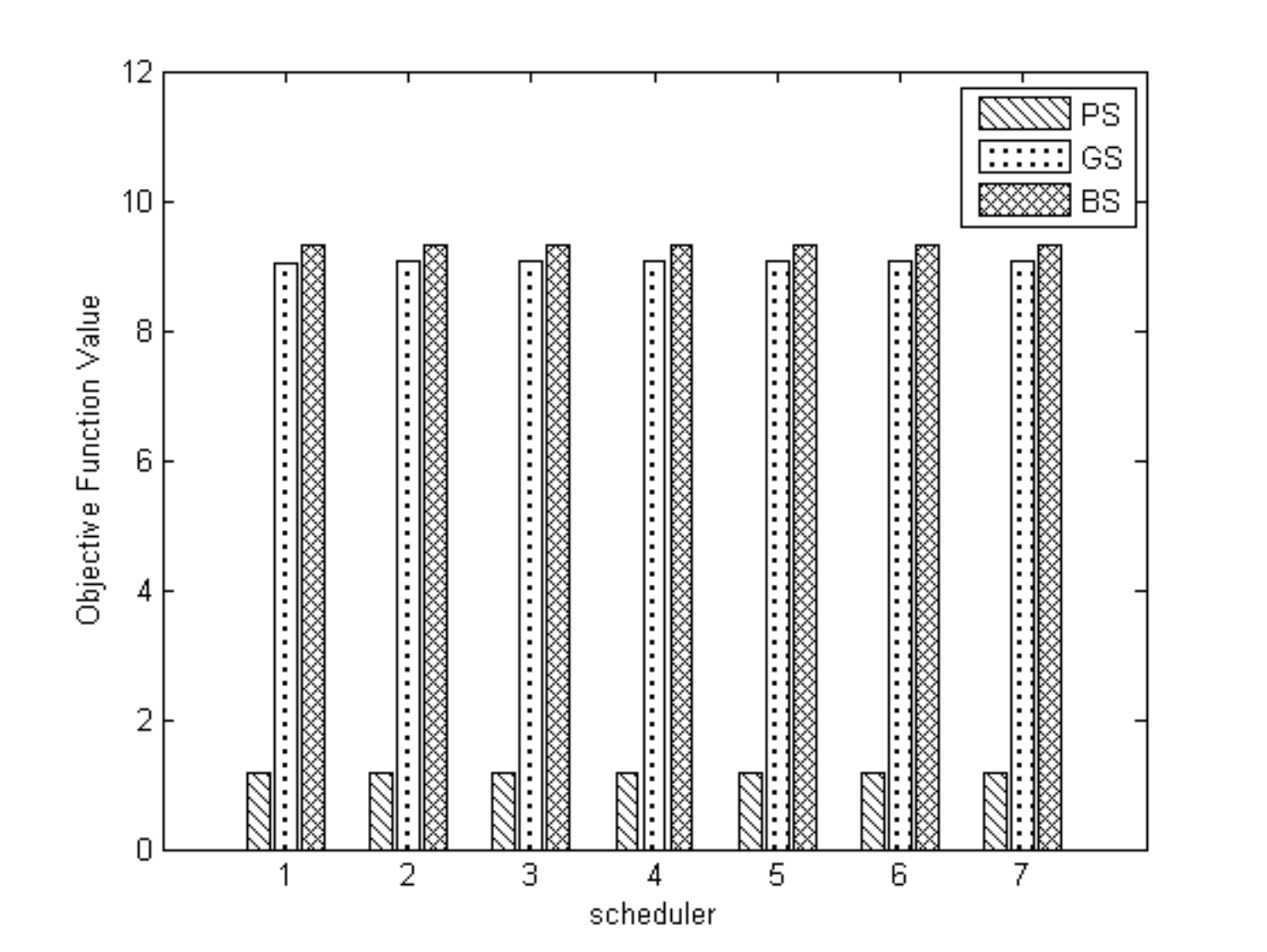}
 \caption{ Each Scheduler's Objective Function Value versus Communication Bandwidth}
 \label{Fig.11}
 \end{center}
\end{figure}

In the previous experiments, we assume that the communication bandwidth is const. In fact, by using the new transmission medium, the communication bandwidth gets faster and faster , which will affect the task scheduling algorithm. In this set of experiment, we will study its effects on the response time and objective function value when the communication bandwidth is $100$, $500$ and $1024$ Kbps. The remaining conditions are the same as the second set of experiment in the first experiment. Fig.\ref{Fig.11}  shows the each scheduler's objective function values when the communication bandwidth changes under different task scheduling algorithm.

From Fig.\ref{Fig.11}, we can see that when the communication bandwidth increases, each scheduler's objective function value decreases; each scheduler's objective function value are little difference under each algorithm; PS is still significantly better than GS and GS is a little better than BS.

\begin{figure}
 \begin{center}
  \includegraphics[width=2.5cm,height=2.5cm]{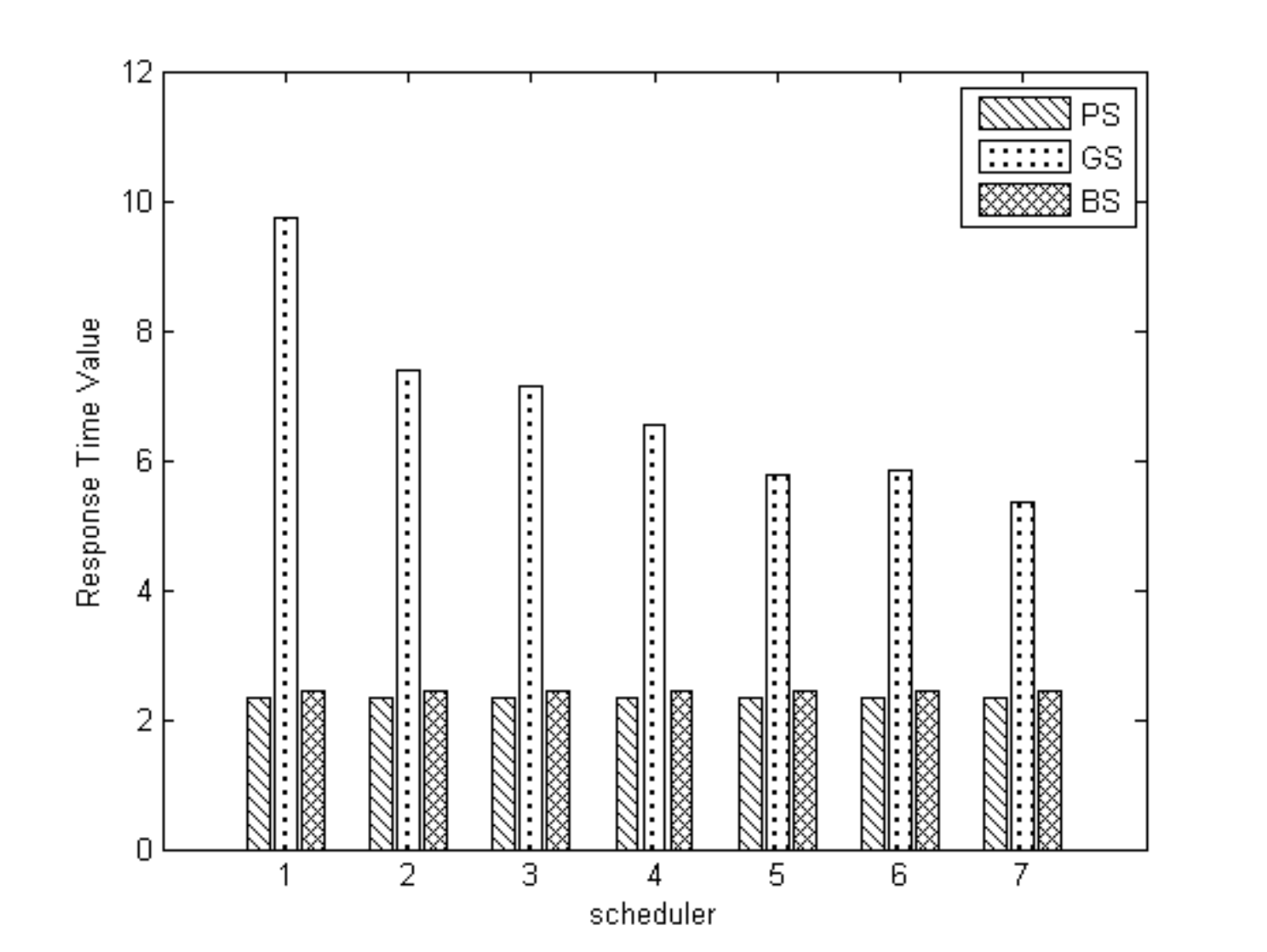}
  \includegraphics[width=2.5cm,height=2.5cm]{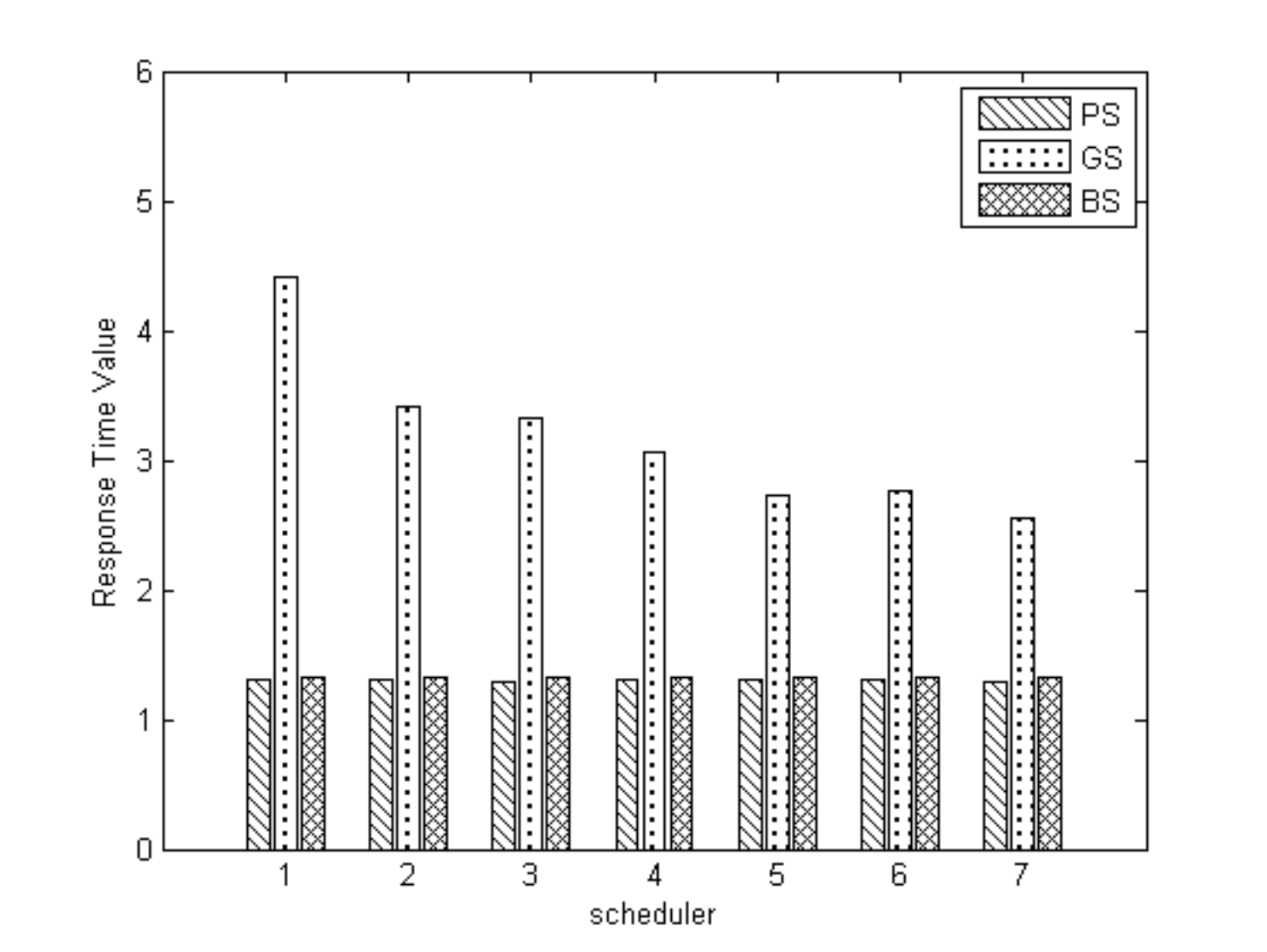}
  \includegraphics[width=2.5cm,height=2.5cm]{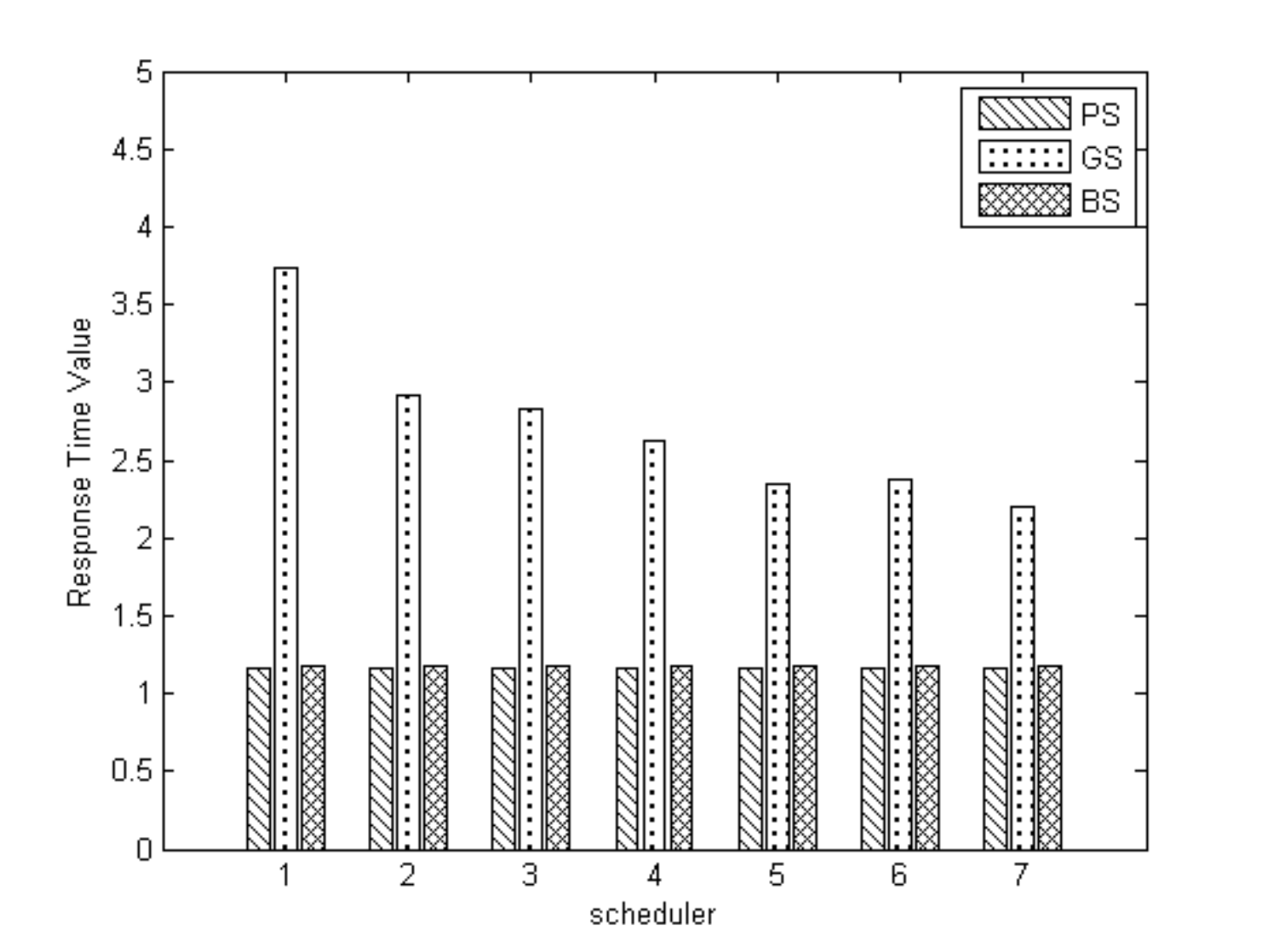}
   \caption{ Each Scheduler's Response Time Value versus Communication Bandwidth}
 \label{Fig.12}
 \end{center}
\end{figure}

Each scheduler's response time values when the communication bandwidth changes under different task scheduling algorithm are shown in Fig.\ref{Fig.12}.


As can be seen from Fig.\ref{Fig.12}, with the communication bandwidth increasing, each scheduler's response time value decrease; each scheduler's response time value has little difference under PS and BS, but for GS.

\subsection{Fairness}

In this part of the experiment, we investigate the fairness of each algorithm. When the average response time value for each scheduler is the same, the fairness is achieved. If one scheduler's response time value is lower and another's is higher, then the scheduling algorithm can be considered unfair. A fairness index\cite{2} is given by
\begin{equation}
  FI=\frac{(\sum_{i=1}^{n}D_{i})^2}{n\sum_{i=1}^{n}D_{i}^{2}}
\end{equation}

Where $D_{i}$ is the average response time of scheduler $i$. When a scheduling algorithm's fairness index is closer to $1.0$, it is more fair. Here, we study the above three scheduling algorithm's fairness under different conditions such as system load, system size. Their initial conditions are the same as the above.

\begin{figure}
 \begin{center}
  \includegraphics[width=8cm,height=8cm]{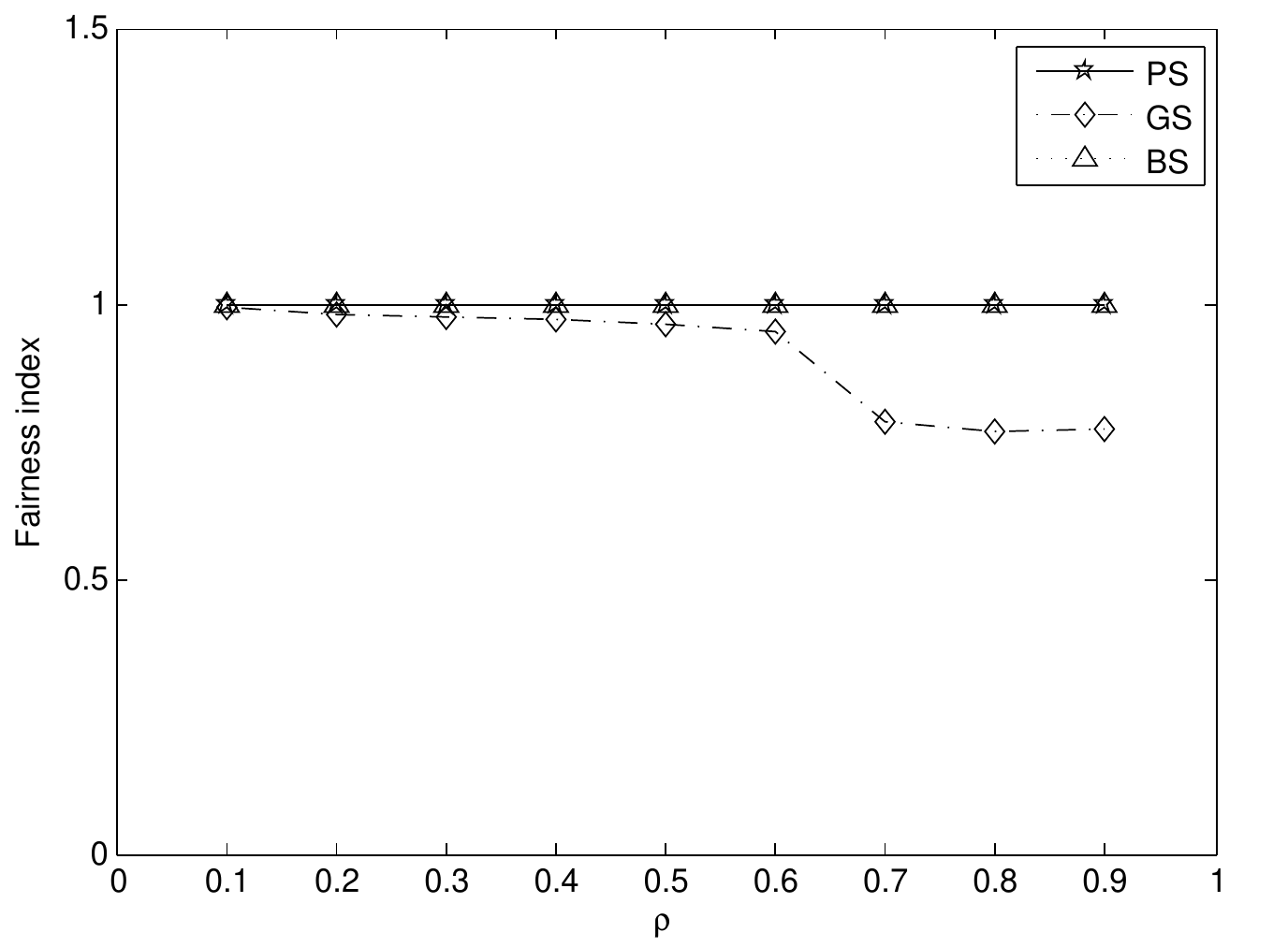}
 \caption{Fairness versus System Load}
 \label{Fig.13}
 \end{center}
\end{figure}

Fig.\ref{Fig.13} shows the fairness index when the system load varies from $0.1$ to $0.9$ under different algorithms. From this figure, we can find that the fairness indexes of both PS and BS are very closer to $1$, which means that PS and BS are fair. But for GS, when the system load is low, the fairness is high. However, the fairness goes lower when the system load gets higher.

\begin{figure}
 \begin{center}
  \includegraphics[width=8cm,height=8cm]{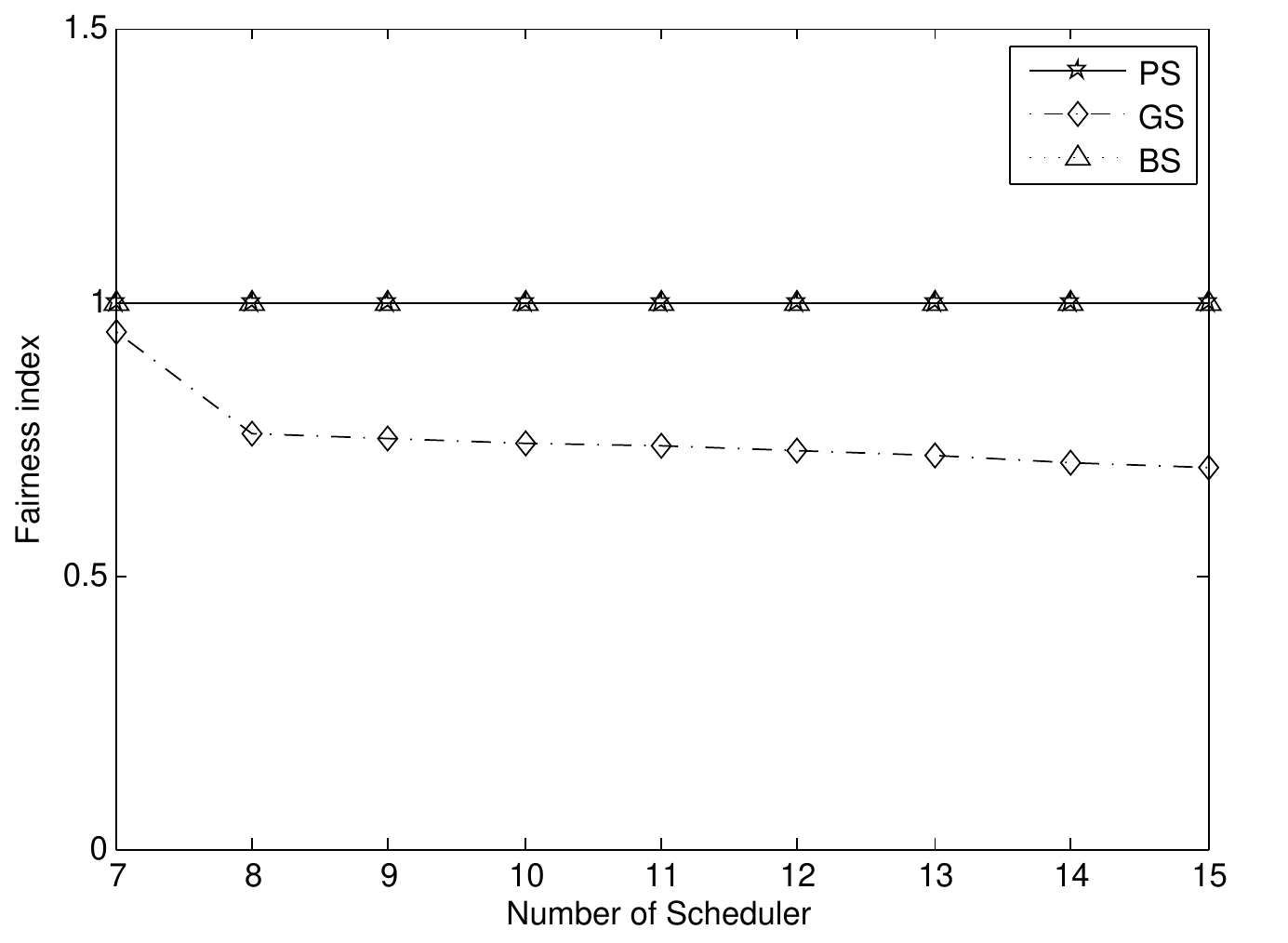}
 \caption{Fairness versus Number of Scheduler}
 \label{Fig.14}
 \end{center}
\end{figure}

\begin{figure}
 \begin{center}
  \includegraphics[width=8cm,height=8cm]{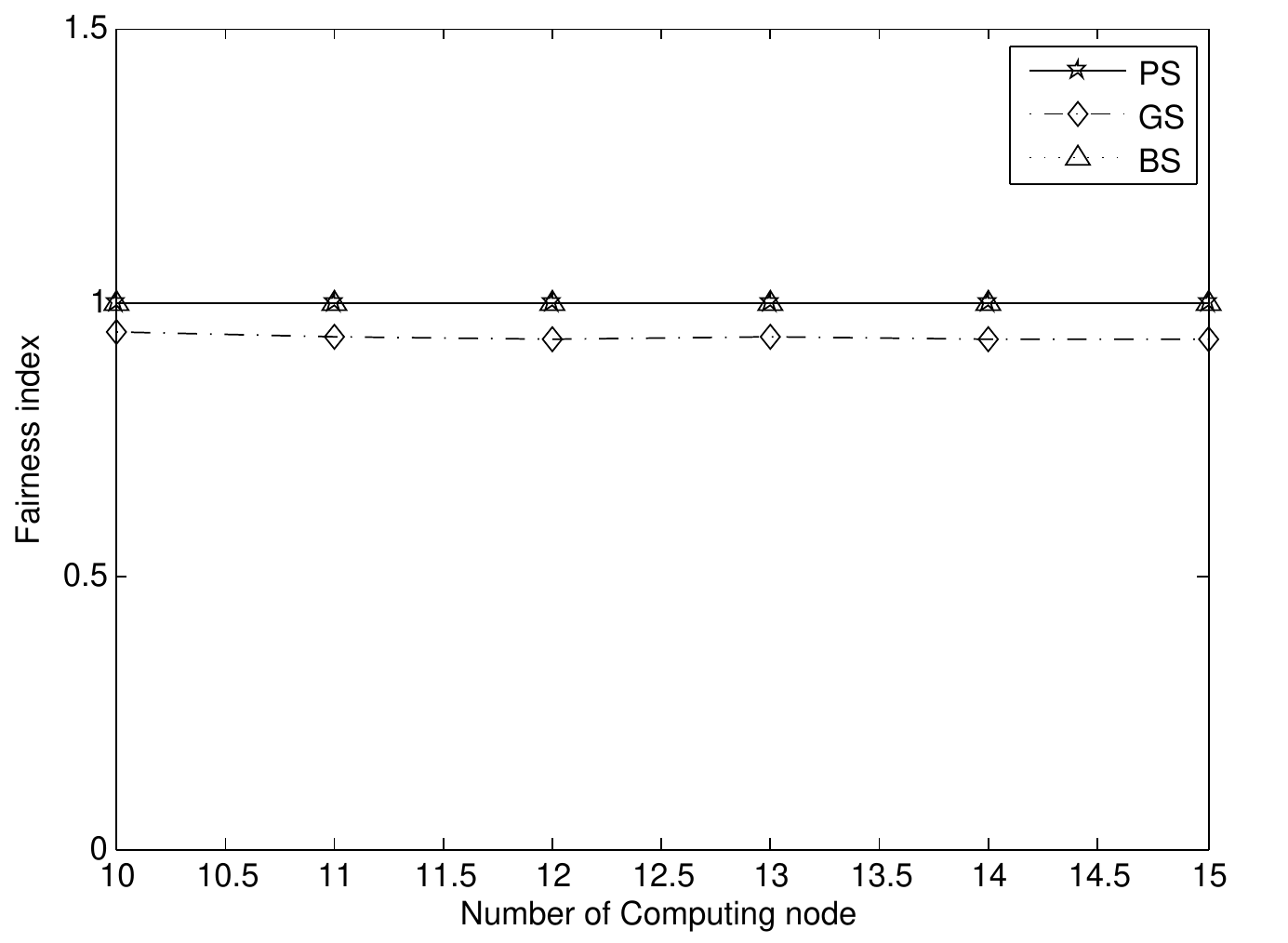}
 \caption{Fairness versus Number of Computing Node}
 \label{Fig.15}
 \end{center}
\end{figure}

When the number of scheduler changes, the fairness indexes of different algorithms are shown in Fig.\ref{Fig.14}. Fig.\ref{Fig.15} shows the fairness indexes when the number of computing node changes. From Fig.\ref{Fig.14} and Fig.\ref{Fig.15}, we can find that the fairness indexes are close to 1.0 all the time for PS and BS; but for GS, when the system size increases, its fairness index decreases.

From the above three experiments , we can draw this conclusion: our proposed algorithm is fair.

\section{Conclusions}

This paper discussed a new task scheduling algorithm about response time. On the premise that a task can be broken down into task slices, we build a mathematical model in which task slices are actually executed in parallel and design a new task allocation schema. From the above experiments, we can draw these conclusions: our schema is fair, stable and has a significantly advantage than the game-theoretic algorithm and balanced scheduling algorithm.

%
%

\label{lastpage}


\begin{thebibliography}{Lam94}

   \bibitem{1}I. FOSTER, Y. ZHAO, I. RAICYI, et al. \emph{Cloud Computing and Grid Computing 360-degree Compared}, In: Proc. of the 2008 Grid Computing Environments Workshop(GCE2008), 1--10.

   \bibitem{2}R. SUBRATA, A. Y. ZOMAYA, B. A. LANDRELDT. \emph{A Cooperative Game FrameWork for QoS Guided Job Allocation Schemes in Grids}, IEEE Transactions on Computers, 2008, 57(10):1413--1422.

   \bibitem{3}C. KIM, H. KAMEDA. \emph{An Algorithm for Optimal Static Load Balancing in Distributed Computer Systems}, IEEE Transactions on computers,1992,41(3):381--384.

   \bibitem{4}D. GROSU, M. LEUNG. \emph{Load Balancing in Distributed Systems: An Approach Using Cooperative Games}, In: Proc. of the International Parallel and Distributed Processing Symposium, 2002,1530--2075/02.

   \bibitem{5}L. MNI, K. HWANG. \emph{Optimal Load Balancing in a Multiple Processing System with Many Job Classes}, IEEE Transactions on software engineering,1985,SE-11(5):491--496.

   \bibitem{6}J. LI, H. KAMEDA. \emph{Load Balancing Problems for Multiclass Jobs in Distributed/Parallel Computer Systems}, IEEE Transactions on computers,1998,47(3):322--332.

   \bibitem{7}A. G. DELAVAR,  M. NEJADKHEIRALLAH, M. MOTALLEB. \emph{A New Scheduling Algorithm for Dynamic Task and Fault Tolerant in Heterogeneous Grid Systems Using Genetic Algorithm}, In: Proc. of the IEEE International conference on computer science and information technology, 2010, 9:408--412.

   \bibitem{8}N. FUJIMOTO, K. HAGIHARA. \emph{Near-optimal Dynamic Task Scheduling of Independent Coarse-grained Tasks onto a Computational Grid}, In: Proc. of the 2003 international conference on parallel processing,2003:391--398.

   \bibitem{9}R. MAHAJAN, M. RODRIG, D. WETHERALL, et al. \emph{Experiences Applying Game Theory to System Design}, In: Proc. of the 2004 Annual Conference of The Special Interest Group on Data Communication [C], Portland, Oregon, USA:ACM Press, 2004:183--190.

   \bibitem{10}K. RANGANATHAN, M. RIPEANU, A. SARIN, et al. \emph{Incentive Mechanisms for Large Collaborative Resource Sharing}, In: Proc. of the 4th IEEE/ACM International Symposium on Cluster Computing and the Grid [C], Chicago, Illinois, USA: IEEE Computer Society, 2004:1--8.

   \bibitem{11}D. GROSU, A. T. CHRONOPOULOS. \emph{Non-cooperative Load Balancing in Distributed Systems}, Journal of Parallel and Distributed Computing, 2005, 65:1022--1034.

   \bibitem{12}K. YI, R. WANG. \emph{Nash Equilibrium Based task Scheduling Algorithm of Multi-schedulers in  Grid Computing}, ACTA Electronica Sinica, 2009, 37(2):329--333.

   \bibitem{13}R. SUBRATA, A. Y. ZOMAYA, B. LANDFELDT. \emph{Game Theoretic Approach for Load Balancing in Computational Grids}, IEEE Transactions on Parallel and Distributed Systems, 2008, 19(1):66--76.

   \bibitem{14}G. WEI, A. V. VASILAKOS, N. XIONG. \emph{Scheduling Parallel Cloud Computing Services: An Evolutional Game}, In: Proc. of the 1st International Conference on Information Science and Engineering(ICISE2009), 2009:376--379.
   \bibitem{15} O.M. Elzeki,  M. Z. Reshad, M. A. Elsoud. \emph{ Improved Max-Min Algorithm in Cloud Computing}, International Journal of Computer Applications, 2012, 50.

   \bibitem{16}G. LIU, J. LI, J. XU. \emph{An Improved Min-min Algorithm in Cloud Computing}, In: Proc. of the 2012 International Conference of Modern Computer Science and Applications, 2013, 191:47--52.

   \bibitem{17}C Zhao, S Zhang, Q Liu. \emph{Independent tasks scheduling based on genetic algorithm in cloud computing}, Wireless Communications, Networking and Mobile Computing, 2009. WiCom'09. 5th International Conference on. IEEE, 2009: 1-4.

   \bibitem{18}K. Li, G. Xu, G. Zhao. \emph{ Cloud Task Scheduling Based on Load Balancing Ant Colony Optimization}, Chinagrid Conference (ChinaGrid), 2011 Sixth Annual. IEEE, 2011: 3-9.

   \bibitem{19}S. S.CHANHAN, R. JOSHI. \emph{A Heuristic for QoS Based Independent Task Scheduling in Grid Environment}, In: Proc. of the International Conference on Industrial and Information system. 2010:102--106.

   \bibitem{20}M. XU, L. CUI, H. WANG, Y. BI. \emph{A Multiple QoS Constrained Scheduling Strategy of Multiple Workflows for Cloud Computing}, In: Proc. of the 2009 IEEE International Symposium on Parallel and Distributed Processing with Application,2009:629--634.

   \bibitem{21}E. DEELMAN, G. SINGH, M. LIVNY, et al£®\emph{The Cost of Doing Science on the Cloud: the Montage Example}, In: Proc. of the ACM/IEEE Conference on Supercomputing£®Piscataway:IEEE Press, 2008:1--12.

   \bibitem{22}M. ASSUNCAO, A. COSTANZO, R. BUYYA. \emph{Evaluating the Cost-benefit of Using Cloud Computing to Extend the Capacity of Clusters}, In: Proc. of the 18th ACM International Symposium on High Performance Distributed Computing£®New York: ACM Press£¬2009:141--150£®

   \bibitem{23}G. TIAN, D. MENG, J. ZHAN. \emph{Reliable Resource Provision Policy for Cloud Computing}, Chinese Journal of Computers. 2010, 33(10):1859--1872.

   \bibitem{24}R. SUN, J. LI. \emph{The Basis of Queue Theory}, Beijing, China: Science Publisher, 2002.

   \bibitem{25}H. Tang,X. Qin. \emph{Practical Optimization Methods}, China: Dalian University of Technology Press. 2005.

   \bibitem{26} Q. Yang. \emph{The Entropy Function Methods For Solving Minimax Problems}, Acta Scientiarum Naturalium Universitatis Nankaiensis, 2001, 34(3):7--15.

   \bibitem{27} Y. CHOW, W. KOHLER. \emph{Models for Dynamic Load Balancing in a Heterogeneous Multiple Processor System}, IEEE Transactions on Computers, 1979, 28:354-361.

\end{thebibliography}
\end{document}